\documentclass[journal]{IEEEtran}

\ifCLASSINFOpdf
\else
\fi
\usepackage{cite}
\usepackage[colorlinks, citecolor=black, linkcolor=black]{hyperref}
\usepackage{times}
\usepackage{epsfig}
\usepackage{graphicx}
\usepackage{amsmath}
\usepackage{amssymb}
\usepackage[capitalize]{cleveref}
\usepackage{color}
\usepackage{multirow}
\usepackage{array}
\usepackage{epstopdf}
\usepackage{subfig}
\usepackage{caption}
\usepackage[font=small]{caption}
\usepackage{booktabs}
\usepackage{enumerate}
\usepackage{float}
\usepackage{arydshln}

\newcounter{MYtempeqncnt}

\newcommand{\etal}{\textit{et al}. }

\newcommand{\tabincell}[2]{\begin{tabular}{@{}#1@{}}#2\end{tabular}}
\newcommand{\eg}{\textit{e.g}.}
\newcommand{\ie}{\textit{i.e}.}
\newcommand{\etc}{\textit{etc}}

\begin{document}
	
\title{Image-specific Convolutional Kernel Modulation for \\Single Image Super-resolution}

\author{Yuanfei~Huang,
	Jie~Li,
	Yanting~Hu,
	Hua~Huang,~\IEEEmembership{Senior Member,~IEEE},
	and~Xinbo~Gao,~\IEEEmembership{Senior Member,~IEEE}
\thanks{
Yuanfei Huang and Hua Huang are with the School of Artificial Intelligence, Beijing Normal University, Beijing, 100875, China. E-mail: \{yfhuang, huahuang\}@bnu.edu.cn.

Jie Li is with the Video and Image Processing System Laboratory, School of Electronic Engineering, Xidian University, Xi'an 710071, China. E-mail: leejie@mail.xidian.edu.cn.

Yanting Hu is with the School of Medical Engineering and Technology, Xinjiang Medical University, Urumqi 830011, China. E-mail: yantinghu2012@gmail.com.

Xinbo Gao is with the Chongqing Key Laboratory of Image Cognition, Chongqing University of Posts and Telecommunications, Chongqing 400065, China (E-mail: gaoxb@cqupt.edu.cn) and with the School of Electronic Engineering, Xidian University, Xi'an 710071, China (E-mail: xbgao@mail.xidian.edu.cn).

({\em Corresponding author: Xinbo Gao})
}}


\maketitle

\begin{abstract}
Recently, deep-learning-based super-resolution methods have achieved excellent performances, but mainly focus on training a single generalized deep network by feeding numerous samples. Yet intuitively, each image has its representation, and is expected to acquire an adaptive model. 
For this issue, we propose a novel image-specific convolutional kernel modulation (IKM) by exploiting the global contextual information of image or feature to generate an attention weight for adaptively modulating the convolutional kernels, which outperforms the vanilla convolution and several existing attention mechanisms while embedding into the state-of-the-art architectures without any additional parameters. 
Particularly, to optimize our IKM in mini-batch training, we introduce an image-specific optimization (IsO) algorithm, which is more effective than the conventional mini-batch SGD optimization.
Furthermore, we investigate the effect of IKM on the state-of-the-art architectures and exploit a new backbone with U-style residual learning and hourglass dense block learning, terms U-Hourglass Dense Network (U-HDN), which is an appropriate architecture to utmost improve the effectiveness of IKM theoretically and experimentally. 
Extensive experiments on single image super-resolution show that the proposed methods achieve superior performances over state-of-the-art methods. Code is available at \href{https://github.com/YuanfeiHuang/IKM}{github.com/YuanfeiHuang/IKM}.
\end{abstract}


\IEEEpeerreviewmaketitle
\section{Introduction}
\IEEEPARstart{S}{ingle} image super-resolution (SR) aims at reconstructing a high-resolution (HR) image from a single low-resolution (LR) image obtained by limited imaging devices, which is considered as a challenging ill-posed inverse problem and widely used in computer vision applications where high-frequency details are greatly desired, such as medical imaging, security, and surveillance.

For decades, numerous methods have been proposed to solve this ill-posed problem. For example, the interpolation-based~\cite{Bicubic1981TASSP,ZhangL2006TIP}, the reconstruction-based~\cite{MarquinaA2008JSC,DongW2011TIP}, and the example-learning-based~\cite{YangJ2010TIP,ZeydeR2010,TimofteR2014ACCV,ZhuZ2014TMM,HuY2016TIP,HuangY2018TIP} methods.
Recently, with the development of high-profile graphic process units (GPUs) and large-scale visual datasets, convolutional neural network (CNN) has obtained continuous attention as a dominant machine learning method in a majority of computer vision applications~\cite{Lecun1998IEEE,Krizhevsky2012NeurIPS,Simonyan2015ICLR,Szegedy2015CVPR,HeK2016CVPR,HuangG2017CVPR,HuJ2018CVPR}.
Moreover, benefited from the excellent ability to end-to-end nonlinear mapping on paired images, deep-learning-based methods have been also exploited for pixel-wise image reconstruction in low-level vision.

\begin{figure}[t]
	\begin{center}
		\includegraphics[width=1\linewidth]{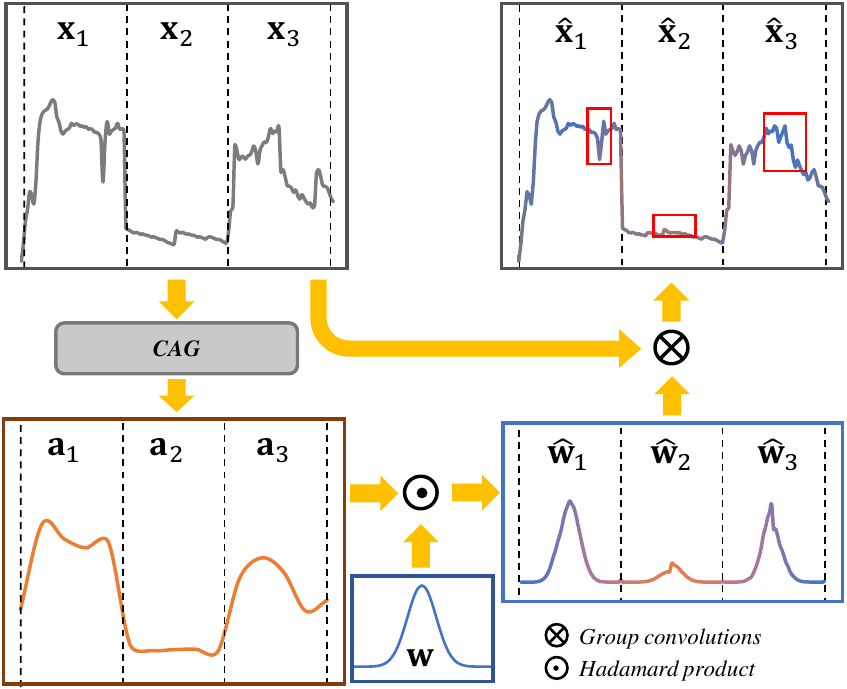}
	\end{center}
	\caption{Illustration of IKM on 1D synthesized data. Particularly, features in each batch could be self-enhanced by adaptively applying the group convolutions with image-specific modulated kernel weights.}
	\label{fig:RefConv_illustration}
\end{figure}
As the first attempt, Dong~\etal\cite{DongC2014ECCV,DongC2016TPAMI} proposed a shallow super-resolution convolutional neural network (SRCNN) by stacking several activated convolutional layers to learn the nonlinear mappings of LR-to-HR pairs, which outperforms most existing example-based SR methods and leads the trend of using end-to-end networks for SR. To facilitate the training of deep CNNs, Kim \etal further proposed the accurate VDSR~\cite{KimJ2016CVPR_VDSR} and firstly demonstrated the effectiveness of residual learning in end-to-end SR networks. However, training a very deep network is still hard as gradient vanishing/exploding, then local residual learning from ResNet~\cite{HeK2016CVPR} was arisen in SR~\cite{LedigC2017CVPR,TaiY2017CVPR,LimB2017CVPRW} and achieved excellent performances. Besides, aiming at fully using the information flow of intermediate features in a network, densely connection~\cite{HuangG2017CVPR} was also developed in SR~\cite{TongT2017ICCV,TaiY2017ICCV,ZhangY2018CVPR,ZhangY2021TPAMI,ZhangX2021TMM}. Besides, to reduce the computational complexities, the transposed convolution (deconvolution)~\cite{DongC2016ECCV} and sub-pixel convolution~\cite{ShiW2016CVPR} were proposed to upscale the images/features and limit the spatial size of LR input to speed up the inference. 
Moreover, the informativeness of features is also an important issue to improve the performance, then attention mechanism was highlighted, such as channel attention~\cite{HuJ2018CVPR,ZhangY2018ECCV}, spatial attention~\cite{WooS2018ECCV,HuY2020TCSVT}, non-local attention~\cite{WangXL2018CVPR,ZhangY2019ICLR}, and task-specific attention~\cite{HuangY2021TIP}.
Nevertheless, these methods mainly focus on training a single generalized deep network by feeding numerous samples, \ie, data-driven method. 

Intuitively, each image has its representation, and then is expected to acquire an adaptive model. To implement this expectation, image-specific SR methods have been proposed by zero-shot learning~\cite{ShocherA2018CVPR,UlyanovD2018CVPR}, but they are time-consuming as training an image-specific model from scratch.
For a similar purpose but to be more efficient, we expect to bridge the gap between the data-driven and image-specific SR methods, and particularly make the following key contributions:
\begin{itemize}
\item We propose a novel image-specific convolutional kernel modulation (IKM) by exploiting the global contextual information of image/feature to generate an attention weight for adaptively modulating the convolutional kernels, which outperforms the vanilla convolution and other attention mechanisms when embedding into the state-of-the-art architectures without any additional parameters. An intuitive illustration of IKM on 1D synthesized data is shown in Fig.~\ref{fig:RefConv_illustration}.

\item We propose an image-specific optimization (IsO) algorithm for optimizing IKM in mini-batch training, which is a feasible algorithm to optimize the kernel weights and backward propagate the image-specific gradients, both of which positively guide the optimization on IKM.

\item We design a new U-hourglass dense network (U-HDN) as the backbone to improve the capacity of model, by utilizing U-style residual leaning and hourglass dense block learning, which are demonstrated to be appropriate architectures for utmost improving the effectiveness of IKM theoretically and experimentally.

\end{itemize}
\begin{figure*}
	\centering
	\includegraphics[width=0.9\linewidth]{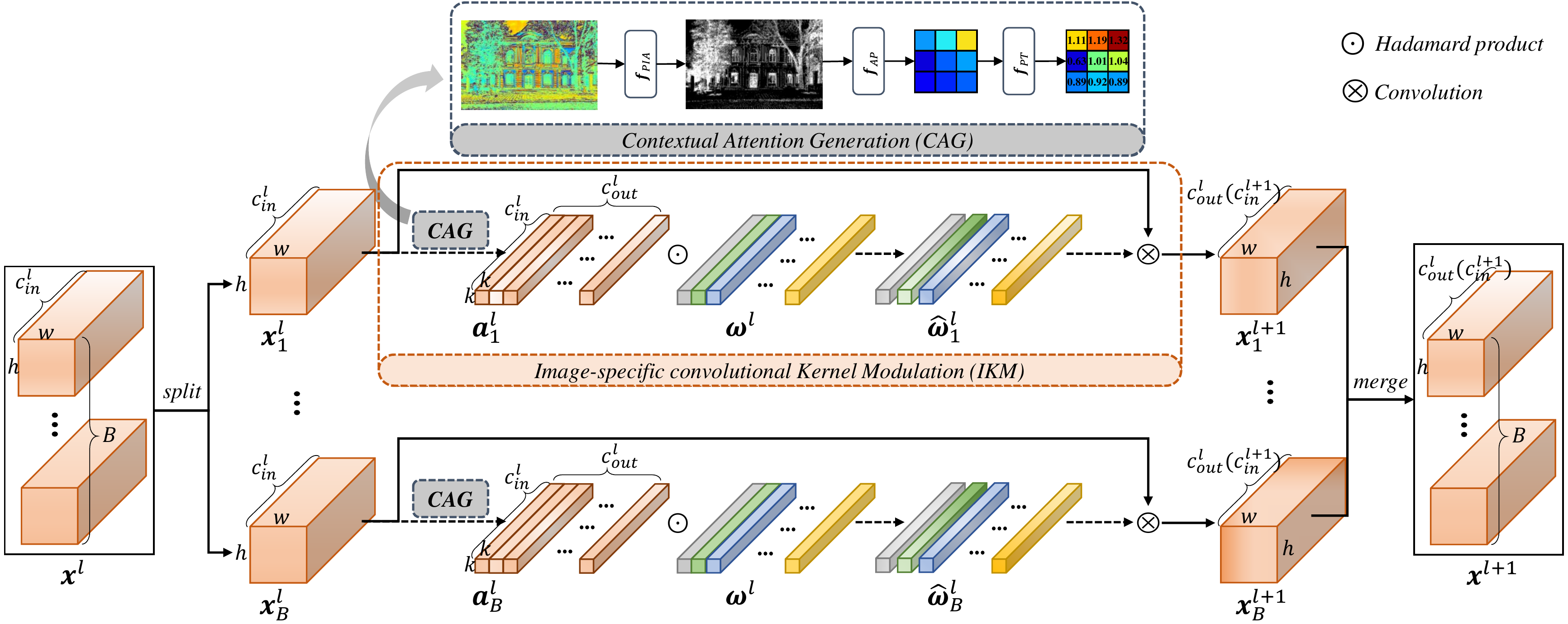}
	\caption{The pipeline of IKM. In $l$-th layer, $\{\boldsymbol{x}^l_i\}_{i=1}^B$ are firstly split in batch-wise into $B$ inputs, where each $\boldsymbol{x}^l_i$ is fed into a contextual attention generation module (CAG) to generate kernel attentions $\boldsymbol{a}_i^l$, and then self-enhanced in local receptive field with the modulated kernels ${\boldsymbol{\hat w}}_i^l$. And finally these enhanced features $\{{\boldsymbol{\hat x}}_i^l\}_{i=1}^B$ are merged into a mini-batch to next layer as $\{\boldsymbol{x}_i^{l+1}\}_{i=1}^B$.}
	\label{fig:architecture of IKM}
\end{figure*}
\section{Related Work}
Since the proposed IKM is implemented by generating an image-specific attention weight to modulate the convolutional kernels, we then describe several attention mechanisms and the deep-learning-based SR methods in this section.
\subsection{Attention Mechanisms}
As the complexity of networks increases, the redundancy of learned features becomes an urgent issue. For efficiently exploiting a scalar response to represent the relative importance of components in feature, some recent researches have applied attention mechanism into the backbone architectures for image classification~\cite{HuJ2018CVPR}, image captioning~\cite{ChenL2017CVPR,XuK2015ICML}, low-level vision~\cite{ZhangY2018ECCV,ZhangY2019ICLR,HuY2020TCSVT,HuangY2021TIP}, and \etc. In particular, the implementation of attention mechanism can be divided into two branches: {\em post hoc network analysis} and {\em trainable attention mechanism}~\cite{JetleyS2018ICLR}. The post hoc network analysis~\cite{JetleyS2018ICLR,ZhangJ2018IJCV} is implemented by back-propagating the excited responses with respect to the high score target, with no need for additional learnable parameters. Moreover, as an end-to-end architecture, the trainable attention mechanism has been widely applied and developed by utilizing the forward feature maps to generate its spatial-attention or channel-attention weights through several additional non-linear transformations and further recalibrating the features in the corresponding locations or channels. Hu~\etal proposed the squeeze-and-excitation networks (SENet)~\cite{HuJ2018CVPR} to recalibrate the features with learnable channel-attention for improving the diversity of features, and has been extended into super-resolution, \eg, residual channel attention network (RCAN)~\cite{ZhangY2018ECCV}. In order to utilize the contextual spatial information in features, two-step attention mechanisms are then raised for gathering the spatial-wise enhancement and channel-wise recalibration, \eg, CBAM~\cite{WooS2018ECCV} and CSFM~\cite{HuY2020TCSVT}. Furthermore, to enable the network to focus on the details, Huang~\etal proposed the interpretable detail-fidelity attention network (DeFiAN)~\cite{HuangY2021TIP} by theoretically demonstrating the representative capacity of Hessian features.

\subsection{Deep-learning-based SR}
Initially, SRCNN~\cite{DongC2016TPAMI} was proposed to learn a nonlinear LR-to-HR mapping function as a high-capacity dictionary for sparse representation. To build deeper networks, Kim \etal further proposed the accurate VDSR~\cite{KimJ2016CVPR_VDSR} and firstly demonstrated the effectiveness of residual learning in end-to-end SR networks. Furthermore, SRResNet~\cite{LedigC2017CVPR} was proposed by straightly stacking the post-activated residual blocks~\cite{HeK2016CVPR} to build a very deep backbone. However, specifically to single image super-resolution, batch normalization (BN) is undesired as weakening the diverse features, then EDSR~\cite{LimB2017CVPRW} was proposed by removing all the BN operations in SRResNet and achieved excellent performances.
Furthermore, inspired by DenseNet~\cite{HuangG2017CVPR}, dense connection was applied into SR tasks by densely integrating the information flow of intermediate features, \eg, SRDenseNet~\cite{TongT2017ICCV}, MemNet~\cite{TaiY2017ICCV}, RDN~\cite{ZhangY2018CVPR,ZhangY2021TPAMI} and GLADSR~\cite{ZhangX2021TMM}.

However, as the depth and width of the network increase, higher computational complexity becomes a tough issue. Hence lightweight architectures are urgently needed for mobile applications with lower storage and lower computational capacities than the large cluster of GPUs. For example, aiming at reducing the space complexity for storage, DRCN~\cite{KimJ2016CVPR_DRCN} and DRRN~\cite{TaiY2017CVPR} stacks deep convolution layers with weight sharing, which recursively calls a single block throughout the whole network. Meanwhile, to reduce the time complexity in model training and testing, lightweight architectures are designed using information diffluence, which splits the features and processes them separately in a manner of connection or convolution, \eg, IDN~\cite{HuiZ2018CVPR}, MSRN~\cite{LiJ2018ECCV}, CARN~\cite{AhnN2018ECCV} and MS$^3$-Conv~\cite{FengR2020ECCV}. Moreover, to deal with this time-consuming issue, the transposed convolution (deconvolution)~\cite{DongC2016ECCV} and sub-pixel convolutions~\cite{ShiW2016CVPR} were developed to upscale the features in the tail, and limit the spatial size of LR input to alleviate the time complexity. Furthermore, for larger upscaling factors (\eg, $\times8$ upscaling), Laplacian pyramid networks were proposed for super-resolution by reconstructing the sub-band residual HR images at multiple pyramid levels~\cite{LaiW2019TPAMI,ZhangD2021TMM}.

\section{Proposed Method}
\label{sec:propose method}
In this section, we first describe the motivation, inference, and optimization of IKM, and the conditions on utmost improving the performance of IKM. Under these conditions, we further describe the U-HDN architecture.

\subsection{Image-specific Convolutional Kernel Modulation}\label{sec:IKM}
To improve the ability to non-linear representation, the existing state-of-the-art SR methods tend to build deeper or wider architectures. Particularly, by adaptively enhancing the features channel-wise or spatially, the attention mechanism has been demonstrated to be effective to improve the performance concerning the specific tasks.

Specifically, conventional deep CNNs generally straightly use convolutional kernels to convolve with the inputs, \ie, in $l$-th layer, the input feature $\{\boldsymbol{x}^l_i,i=1,2,...,c^l_{in}\}$ would be convolved with the kernel weights $\boldsymbol{w}^l$ to get the output feature $\{\boldsymbol{y}^{l}_j,j=1,2,...,c^l_{out}\}$ as
\begin{equation}
\boldsymbol{y}_{j}^{l}=\sum\limits_{i=1}^{c^l_{in}}{\boldsymbol{x}_{i}^{l}\otimes \boldsymbol{w}_{j,i}^{l}}+b_{j}
\label{eq:conventional conv}
\end{equation}
where, $c_{in}^l$ and $c_{out}^l$ denote the number of input and output channels in $l$-th layer.
Specifically, in the local receptive field $\mathcal{R}$, for each location $p_0$ on the output feature $\boldsymbol{y}^{l}_j$, we have
\begin{equation}
\boldsymbol{y}_{j}^{l}({{p}_{0}})=\sum\limits_{i=1}^{c^l_{in}}{\sum\limits_{p\in \mathcal{R} }{\boldsymbol{x}_{i}^{l}({{p}_{0}}+p)\cdot \boldsymbol{w}_{j,i}^{l}(p)+{{b}_{j}}}}
\label{eq:conventional conv detail}
\end{equation}
Intuitively, there are three ways to enhance the output features $\{\boldsymbol{y}^{l}_j,j=1,2,...,c^l_{out}\}$:

{\em i) recalibration on channel-wise dimension $j$.

ii) recalibration on spatial-wise pixels ${p}_0$.

iii) recalibration on both of them.}

So far, in order to enhance the features with different attentions, the attention mechanisms are introduced into the backbone architecture channel-wise or spatially, which aims at enhancing the output features through recalibrating features in channel-wise dimension and straightly generating attention maps on features, respectively.

\subsubsection{Channel attention}
In the first way, the output features are recalibrated on the channel-wise dimension as
\begin{equation}
\begin{aligned}
\boldsymbol{\hat{y}}_{j}^{l}({{p}_{0}})&={{a}^l_{j}}\boldsymbol{y}_{j}^{l}({{p}_{0}})\\
&={{a}^l_{j}}(\sum\limits_{i=1}^{c_{in}^{l}}{\sum\limits_{p\in \mathcal{R} }{\boldsymbol{x}_{i}^{l}({{p}_{0}}+p)\cdot \boldsymbol{w}_{j,i}^{l}(p))+{{b}_{j}}}}
\end{aligned}
\label{eq:channel attention}
\end{equation}
where, $\{a^l_j, j=1,2,...,c^l_{out}\}$ denotes the channel attention in 1D vector, which is inferred from the initial output features $\boldsymbol{y}_{j}^{l}$. As in SENet~\cite{HuJ2018CVPR},
\begin{equation}
a^l=\boldsymbol{f}_{Sig}(\boldsymbol{f}_{FC}(\boldsymbol{f}_{ReLU}(\boldsymbol{f}_{FC}(\boldsymbol{f}_{AP}(\boldsymbol{y}^{l};K)))))
\label{eq:CA_parametrization}
\end{equation}
where $\boldsymbol{f}_{Sig}$, $\boldsymbol{f}_{FC}$, $\boldsymbol{f}_{ReLU}$ and $\boldsymbol{f}_{AP}(*;K)$ denote the sigmoid function, full connection layer, ReLU activation and average pooling operation with output size of $K=1$, respectively.
\subsubsection{Spatial attention}
In the second way, we should recalibrate the output feature on isolated pixels, so the attention could be represented in 2D maps, and the output features are transformed as
\begin{equation}
\begin{aligned}
\boldsymbol{\hat{y}}_{j}^{l}({{p}_{0}})&={\boldsymbol{a}^l_{j}}\boldsymbol{y}_{j}^{l}({{p}_{0}})\\
&={\boldsymbol{a}^l_{j}}(\sum\limits_{i=1}^{c_{in}^{l}}{\sum\limits_{p\in \mathcal{R} }{\boldsymbol{x}_{i}^{l}({{p}_{0}}+p)\cdot \boldsymbol{w}_{j,i}^{l}(p))+{{b}_{j}}}}
\end{aligned}
\label{eq:spatial attention}
\end{equation}
where, $\{\boldsymbol{a}^l_j, j=1,2,...,c^l_{out}\}$ denotes the spatial attention in 2D matrix at size of output $\boldsymbol{y}$. As in CBAM~\cite{WooS2018ECCV},
\begin{equation}
\boldsymbol{a}^l=\boldsymbol{f}_{Sig}(\boldsymbol{f}_{Conv}([\boldsymbol{y}^l_{mean}, \boldsymbol{y}^l_{max}]))
\label{eq:SA_parametrization}
\end{equation}
where $\boldsymbol{f}_{Conv}$ denotes the 2D convolutions, $\boldsymbol{y}^l_{mean}$ and $\boldsymbol{y}^l_{max}$ respectively represent the mean and maximum maps in channel-wise.

\subsubsection{IKM}
As both channel attention and spatial attention essentially improve the performance by enhancing the informativeness of features, but also suffer the same dilemma of data-driven SR methods, it is urgent to find an effective and efficient way to build an image-specific adaptive model.

For this issue, we revisit the operation of convolution as Eq.~(\ref{eq:conventional conv detail}), and find if we fix the input features $\{\boldsymbol{x}^l_i,i=1,2,...,c^l_{in}\}$, the output features $\{\boldsymbol{y}^{l}_j,j=1,2,...,c^l_{out}\}$ are highly related to the kernels, specifically, the channel-wise dimension $c^l_{in}$ and the size of receptive fields $\mathcal{R}$. Namely, it is feasible to integrate the channel and spatial attention to effectively modulate the kernels.

Therefore, as a new attention mechanism, we propose an image-specific way by utilizing the contextual information of features to modulate the convolutional kernels and re-process it with the modulated kernels, terms Image-specific convolutional Kernel Modulation (IKM) as illustrated in Fig.~\ref{fig:architecture of IKM} and formulated as
\begin{equation}
\boldsymbol{\hat y}_{j}^{l}({{p}_{0}})=\sum\limits_{i=1}^{c_{in}^{l}}{\sum\limits_{p\in \mathcal{R} }{\boldsymbol{x}_{i}^{l}({{p}_{0}}+p)\cdot (\boldsymbol{a}_{j,i}^{l}(p)\cdot\boldsymbol{w}_{j,i}^{l}(p))+{{b}_{j}}}}
\label{eq:reflective convolution_1}
\end{equation}
where, $\{\boldsymbol{a}^l_{j,i}, i=1,2,...,c^l_{in},j=1,2,...,c^l_{out}\}$ denotes the kernel attention weight, and is generated by exploiting the global contextual information.

However, since the kernel attention weights are not immediately generated in the forward inference, namely, the kernel weights are fixed when the network is trained over, it is difficult to optimize $\boldsymbol{a}$ by calculating the gradients of kernel weights $\boldsymbol{w}$. For this issue, we transform Eq.~(\ref{eq:reflective convolution_1}) into
\begin{equation}
\boldsymbol{\hat y}_{j}^{l}({{p}_{0}})=\sum\limits_{i=1}^{c_{in}^{l}}{\sum\limits_{p\in \Re }{(\boldsymbol{a}_{j,i}^{l}(p)\cdot \boldsymbol{x}_{i}^{l}({{p}_{0}}+p))\cdot \boldsymbol{w}_{j,i}^{l}(p)+{{b}_{j}}}}
\label{eq:reflective convolution}
\end{equation}
in this way, we can infer the enhanced output feature $\boldsymbol{\hat y}$ from a group of self-enhanced input features in the local receptive field and input channels, by utilizing a group of attention weights $\boldsymbol{a}\in \mathbb{R}^{c_{out}\times c_{in}\times \mathcal{R}}$. Since the attention maps are considered as acting on the local receptive fields of input features, the optimization of attentions would be conducted on only the input features and the inference of attention is also in a forward processing way.

\begin{figure}[t]
	\centering
	\includegraphics[width=0.9\linewidth]{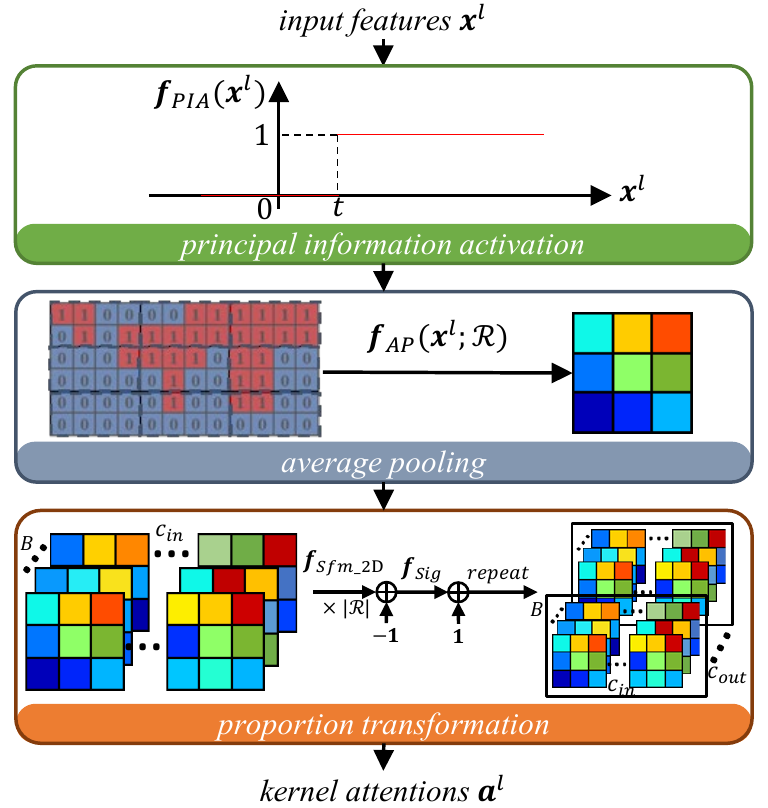}
	\caption{The pipeline of contextual attention generation module $\boldsymbol{f}_{CAG}$ in IKM, which is designed to generate kernel attention weights by inferring the global contextual information of feature map.}
	\label{fig:architecture of CAG}
\end{figure}
Different from the channel attention and spatial attention, IKM needs a 2D matrix of the same size as the kernel. Thus we generate the attention weights $\{\boldsymbol{a}_{j,i}\}$ by using a Contextual Attention Generation module (CAG) $\boldsymbol{f}_{CAG}(\cdot)$, as
\begin{equation}
\boldsymbol{a}^{l}=\boldsymbol{f}^l_{CAG}(\boldsymbol{x}^{l})=\boldsymbol{f}^l_{PT}(\boldsymbol{f}^l_{AP}(\boldsymbol{f}^l_{PIA}(\boldsymbol{x}^{l});\mathcal{R} ))
\label{eq:attention generation}
\end{equation}
where, as illustrated in Fig.~\ref{fig:architecture of CAG}, CAG consists of three parts for different purposes, \ie,

{\em i) Principal information activation $\boldsymbol{f}^l_{PIA}$}. $\boldsymbol{f}^l_{PIA}$ is utilized to sample the dominating pixels to represent the principal contextual information. Specifically, we exploit a hard threshold strategy for principal information marking, which is formulated as
\begin{equation}
\boldsymbol{f}^l_{PIA}(\boldsymbol{x}_{i}^{l})=\left\{\begin{matrix}
1 & ,\boldsymbol{x}_{i}^{l}\ge t\\
0 & ,\boldsymbol{x}_{i}^{l}< t
\end{matrix}\right.
\label{eq:PIA}
\end{equation}
where $t$ is a predefined threshold to sample the dominating pixels, and we assign 1 on them as the principal information.

{\em ii) Average pooling $\boldsymbol{f}^l_{AP}$}. Since the principal information is activated by $\boldsymbol{f}^l_{PIA}$, we then apply the average pooling operation $\boldsymbol{f}^l_{AP}(\boldsymbol{f}^l_{PIA}(\boldsymbol{x}^l);\mathcal{R})$ on the principal information map with output size of $\mathcal{R}$. In a sense, $\boldsymbol{f}^l_{AP}$ here is exploited to calculate the proportion of the dominating pixels in each local patch as in Fig.~\ref{fig:architecture of CAG}.

{\em iii) Proportion transformation $\boldsymbol{f}^l_{PT}$}. As Eq.~(\ref{eq:reflective convolution}), we need a group of attention weights $\boldsymbol{a}\in \mathbb{R}^{c_{out}\times c_{in}\times \mathcal{R}}$, then there are two conditions that need meet: (a) the attention weights should be at the size of ${c_{out}\times c_{in}\times \mathcal{R}}$; (b) the attention weights should be adapted into reasonable values to avoid gradient exploding. For these conditions, we then introduce the proportion transformation module $\boldsymbol{f}^l_{PT}$, which can be formulated as
\begin{equation}
\boldsymbol{a}^l = \boldsymbol{f}^l_{PT}(\boldsymbol{\hat x}^{l})=1+\boldsymbol{f}^l_{Sig}({|\mathcal{R}|}\boldsymbol{f}^l_{Sfm\_2D}(\boldsymbol{\hat x}^{l})-1)
\label{eq:PT}
\end{equation}
where $\boldsymbol{\hat x}=\boldsymbol{f}^l_{AP}(\boldsymbol{f}^l_{PIA}(\boldsymbol{x}^{l});\mathcal{R})$. To meet the condition (b), we utilize the 2D Softmax function $\boldsymbol{f}^l_{Sfm\_2D}$ and Sigmoid transformation function $\boldsymbol{f}^l_{Sig}$ to generate reasonable values of attention. Furthermore, to meet condition (a), we repeat $\boldsymbol{f}^l_{PT}(\mathbf{\hat x}^{l})$ for $c^l_{out}$ times to match the size of $\{\boldsymbol{w}^l_{j,i}\}$, namely, $\boldsymbol{a}_{1,i}^{l}=\boldsymbol{a}_{2,i}^{l}=...=\boldsymbol{a}_{{{c}_{out}},i}^{l}$. Note that, as Eq.~(\ref{eq:PT}), $\boldsymbol{f}^l_{AP}$ is acting in a residual learning fashion, which is more feasible for optimization empirically.

Compared with the existing channel and spatial attention mechanisms, our IKM utilizes the global contextual information to generate a group of attention weights $\boldsymbol{a}$, and adaptively modulates the kernels for feature self-enhancement in local receptive field and input channels. In a sense, IKM is equipped with the superiorities of both spatial and channel attention.
\begin{table}
	\begin{tabular}{p{8.42cm}}
		\toprule
		\normalsize
		\setcounter{MYtempeqncnt}{\value{equation}}
		\setcounter{equation}{11}
		{\bf Algorithm 1} {\em IsO Algorithm in Mini-Batch Training}\\
		\midrule
		{\tt // Image-specific Optimization (IsO) on $l$-th layer}\\
		{\bf\em Input}: $B$ features/images in a mini-batch $\{\boldsymbol{x}^l_1,\boldsymbol{x}^l_2,...,\boldsymbol{x}^l_B\}$, general convolutional weights $\boldsymbol{w}^l$, contextual attention generation module $\boldsymbol{f}^l_{CAG}$;\\
		{\bf\em Output}: optimized general convolutional weights $\boldsymbol{w}^l$, output features/images $\{\boldsymbol{y}^l_1,\boldsymbol{y}^l_2,...,\boldsymbol{y}^l_B\}$.\\
		{\tt // Forward propagation}\\
		\parbox{\linewidth}{
			\begin{enumerate}[1.]
				\item Generating image-specific attentions $\boldsymbol{a}^l$ as Eq.(\ref{eq:attention generation});
				\item Repeat $\boldsymbol{a}^l$ as $[\boldsymbol{a}^l,\boldsymbol{a}^l,...,\boldsymbol{a}^l]_{c^l_{out}}$ of size $(B\times c^l_{out})\times c^l_{in}\times \mathcal{R}^l$;
				\item Repeat $\boldsymbol{w}^l$ as $[\boldsymbol{w}^l,\boldsymbol{w}^l,...,\boldsymbol{w}^l]_B$ of size $(B\times c^l_{out})\times c^l_{in}\times \mathcal{R}^l$;
				\item Modulate kernel weights as
				\begin{equation}
					\boldsymbol{\hat w}^l = \boldsymbol{a}^l \cdot \boldsymbol{w}^l
					\label{eq:recalibrate_kernel}
				\end{equation}
				and $\boldsymbol{\hat w}^l$ is of size $(B\times c^l_{out})\times c^l_{in}\times \mathcal{R}^l$;
				\item Output features/images using group convolutions as
				\begin{equation}
					\boldsymbol{y}^l = GroupConv(\boldsymbol{x}^l;\boldsymbol{\hat w}^l, groups=B)
				\end{equation}
		\end{enumerate}}
		\\
		{\tt // Backward propagation}\\
		\parbox{\linewidth}{
			\begin{enumerate}[1.]
				\item Under the guidance of objective function $\boldsymbol{\mathcal{L}}$, the gradient in $l$-th layer should be
				\begin{equation}
					g^l=\frac{\partial \boldsymbol{\mathcal{L}}}{\partial \boldsymbol{x}^l}\frac{\partial \boldsymbol{x}^l}{\partial {\boldsymbol{\hat w}}^l}
				\end{equation}
				but it is infeasible to use $\frac{\partial \boldsymbol{x}^l}{\partial {\boldsymbol{\hat w}}^l} \in \mathbb{R}^{(B\times c^l_{out})\times c^l_{in}\times \mathcal{R}^l}$ to optimize the parameter ${\boldsymbol{w}}^l\in \mathbb{R}^{c^l_{out}\times c^l_{in}\times \mathcal{R}^l}$. Then
				\item Unfold $\frac{\partial \boldsymbol{x}^l}{\partial {\boldsymbol{\hat w}}^l}$ as $\frac{\partial \boldsymbol{x}^l}{\partial {\boldsymbol{\hat w}}^l} \in \mathbb{R}^{B\times c^l_{out}\times c^l_{in}\times \mathcal{R}^l}$;
				\item Calculate the gradients as
				\begin{equation}
					g^l=\frac{\partial \boldsymbol{\mathcal{L}}}{\partial \boldsymbol{x}^l}\sum_{b=1}^B (\frac{\partial \boldsymbol{x}^l}{\partial {\boldsymbol{\hat w}}^l})_b
				\end{equation}
				\item Optimize the parameters $\boldsymbol{w}^l$ using $g^l$.
		\end{enumerate}}
		\\
		\bottomrule
		\setcounter{equation}{\value{MYtempeqncnt}}
	\end{tabular}
	\vspace{-0.2cm}
	\label{al: algorithm1}
\end{table}
\setcounter{MYtempeqncnt}{\value{equation}}
\setcounter{equation}{15}
\subsection{Image-specific optimization}
As aforementioned, each image should generate its specific contextual attention to adaptively modulate the kernels. However, in mini-batch training, the kernel weights $\{\boldsymbol{w}^l\}$ is of size $c^l_{out}\times c^l_{in}\times \mathcal{R}^l$ and independent of batch-wise dimension, but the attention $\{\boldsymbol{a}^l\}$ from Eq.~(\ref{eq:PT}) is of size $B\times c^l_{out}\times c^l_{in}\times \mathcal{R}^l$ where $B$ denotes the batch size.Thus, it is infeasible to backward the gradient in general mini-batch optimization, \eg, stochastic gradient descent (SGD). 
Therefore, in Algorithm~1, we propose an image-specific optimization (IsO) algorithm for mini-batch training, which is feasible to implement the optimization of modulated kernels and effectively backward propagate the image-specific attention weights, both of which positively guide the optimization of our IKM model.

%
\begin{figure*}
	\centering
	\includegraphics[width=0.92\linewidth]{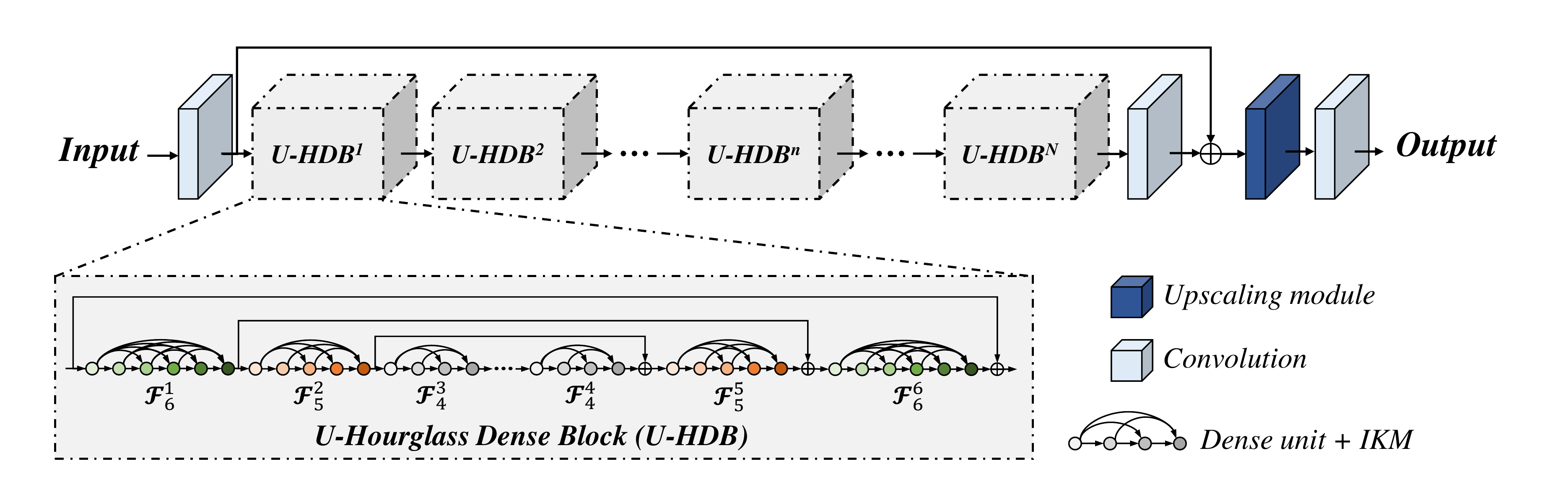}
	\caption{The framework of the proposed U-HDN backbone, which stacks a chain of U-hourglass dense blocks (U-HDB).}
	\label{fig:architecture of UHDN}
\end{figure*}

\subsection{U-Hourglass Dense Network}
\label{sec:UHDN}
Since the attentions are repeated $c_{out}$ times to expand into the size of the kernel, the output channel should be lower to avoid reducing the diversity of attentions. Therefore, in this section, we mainly introduce an effective U-hourglass dense network (U-HDN) to utmost improve the effectiveness of IKM.

As a solution, dense connection has been demonstrated to be an excellent strategy to improve the information flow between layers with concatenated features, as in DenseNet~\cite{HuangG2017CVPR}:
\begin{equation}
{{\boldsymbol{x}}^{l+1}}=\boldsymbol{f}^l_{Composite}([{{\boldsymbol{x}}^{0}},{{\boldsymbol{x}}^{1}},...,{{\boldsymbol{x}}^{l}}])
\label{eq:composite function}
\end{equation}
where $\boldsymbol{f}^l_{Composite}(\cdot)$ indicates the $l$-th composite function which consists of a $3\times3$ convolution and ReLU. Since the number of parameters increases by a large margin with the accumulated inputs, the growth rate $g$ is then applied to reduce the dimension of feature maps. Besides, a transition layer $\boldsymbol{f}_{Transition}(\cdot)$ is applied in the tail of dense unit for reducing the accumulated information flows to a single tensor $y$ with $c$ channels by a $1\times1$ convolution layer without ReLU:
\begin{equation}
{\boldsymbol{y}}=\boldsymbol{f}_{Transition}([{{\boldsymbol{x}}^{0}},{{\boldsymbol{x}}^{1}},...,{{\boldsymbol{x}}^{l}},{{\boldsymbol{x}}^{l+1}}])
\label{eq:transition layer}
\end{equation}

As considered that the dimension in output channel $c_{out}$ should be lower to avoid reducing the diversity of attention, it is better to rich the input features and reduce the output features for more informative attention. Therefore, the dense unit is expected to be an appropriate architecture to improve the effectiveness of IKM.

Furthermore, to improve the effectiveness of dense connection, we then design a U-hourglass dense network (U-HDN) by stacking $N$ U-hourglass dense blocks as illustrated in Fig.~\ref{fig:architecture of UHDN}. Particularly, different from the conventional dense blocks in DenseNet~\cite{HuangG2017CVPR}, we design the U-hourglass dense block (U-HDB) with two major amelioration:

\subsubsection{U-style residual learning}
Residual learning~\cite{HeK2016CVPR} has been widely applied to facilitate training very deep networks, and has been demonstrated to be effective with dense connection by building the residual dense block (RDB)~\cite{ZhangY2018CVPR}. However, from Eq.(\ref{eq:transition layer}), information flow in each layer is fed into the transition layer as identity, then
\begin{equation}
\begin{aligned}
{\boldsymbol{y}_j}&=\sum\limits_{i=1}^{c+(l+1)\cdot g}{{{\boldsymbol{x}}_{i}}\cdot {{\boldsymbol{w}}_{j,i}}}
=\sum\limits_{i=1}^{c}{{{\boldsymbol{x}}_{i}}\cdot {{\boldsymbol{w}}_{j,i}}}+\sum\limits_{i=c+1}^{c+(l+1)\cdot g}{{{\boldsymbol{x}}_{i}}\cdot {{\boldsymbol{w}}_{j,i}}}\\
&={\boldsymbol{x}^0}\otimes{\boldsymbol{w}_{j,1:c}} + [{\boldsymbol{x}^1}, {\boldsymbol{x}^2}, ..., {\boldsymbol{x}^{l+1}}]\otimes{\boldsymbol{w}_{j,c+1:c+(l+1)\cdot g}}
\end{aligned}
\label{eq:U residual learning}
\end{equation}
where $\{\boldsymbol{x}_i\}$ indicates the accumulated information flows and is formulated as $\{\boldsymbol{x}_i\}=[\boldsymbol{x}^0_i,\boldsymbol{x}^1_i,...,\boldsymbol{x}^{l+1}_i]$, $\{\boldsymbol{w}_{j,i}\}$ denotes $1\times1$ kernel weight of transition layer $\boldsymbol{f}_{Transition}$.

Then, the vanilla input feature $\boldsymbol{x}^0$ is straightly propagated as a part of output without any non-linear transformation, in a sense, every single dense unit could be seen as a single layer on part of channels.
Then if we directly introduce the residual learning for a single dense unit as residual dense block (RDB) (as illustrated in Fig.~\ref{fig:UHDB}), the ability of non-linear representation in the dense unit would be largely discounted.

Therefore, we suggest each residual branch has to stack more than two dense units for powerful non-linear representation, and design a more effective U-style residual learning by stacking two symmetric dense units as in Fig.~\ref{fig:architecture of UHDN} and Fig.~\ref{fig:UHDB}.

\subsubsection{Hourglass dense block learning}
As illustrated in Fig.~\ref{fig:architecture of UHDN}, we utilize an hourglass structure in the residual branch. We denote each dense unit by $\boldsymbol{\mathcal{F}}^m_{d_m}(\cdot)$, where $m$ and $d_m$ represent the $m$-th dense unit and the corresponding depth respectively, then the feed-forward procedure could be formulated as
\begin{equation}
\boldsymbol{y}=\boldsymbol{x}+\boldsymbol{\mathcal{F}}_{{{d}_{M}}}^{M}(\boldsymbol{\mathcal{F}}_{{{d}_{1}}}^{1}(\boldsymbol{x})+\boldsymbol{\mathcal{F}}_{{{d}_{M-1}}}^{M-1}(\boldsymbol{\mathcal{F}}_{{{d}_{2}}}^{2}(\boldsymbol{\mathcal{F}}_{{{d}_{1}}}^{1}(\boldsymbol{x}))+...))
\label{eq:hourglass dense block}
\end{equation}
in this case, the information flow of head unit $\boldsymbol{\mathcal{F}}_{{{d}_{1}}}^{1}(\boldsymbol{x})$ is transmitted straightly into tail unit $\boldsymbol{\mathcal{F}}_{{{d}_{M}}}^{M}$ through the residual branch, and also partly into the other intermediate units through dense connection as illustrated in Eq.~(\ref{eq:U residual learning}). Besides, intuitively, the whole information flow of the tail unit works as the residues in such residual learning.
In a sense, for a U-HDB with $M$ dense units, the desired residual features largely depend on the head unit $\boldsymbol{\mathcal{F}}_{{{d}_{1}}}^{1}$ and the tail unit $\boldsymbol{\mathcal{F}}_{{{d}_{M}}}^{M}$.
Moreover, in Eq.(\ref{eq:transition layer}), the deeper dense unit, the more accumulated features to make more informative outputs.

Therefore, deeper head and tail units are of great advantages to generate informative residues. Then, it is feasible to compress the computational complexities in the intermediate parts of U-HDB to get improvement with relatively fewer parameters.
Therefore, as in Fig.~\ref{fig:architecture of UHDN}, we set $d_{M-i+1}=d_i$ and $d_j>d_i$ for $i=1,2,...,M/2, j<i$ in hourglass dense block learning, which is symmetric as an hourglass.

\begin{table*}
	\renewcommand\arraystretch{1.05}
	\centering
	\captionsetup{justification=centering}
	\caption{\textsc{\\Quantitative comparisons with the state-of-the-art lightweight (\#Params$\le$2.0M or \#FLOPs$\le$100G) SISR methods.}}
	\begin{tabular}{c l c c c c c c c c c c c c}
		\hline
		\hline
		\multirow{2}{*}{Scale}&\multirow{2}{*}{Method}&\#Params&\#FLOPs&{Set5}&{Set14}&{BSD100}&{Urban100}&{Manga109}\\
		&&(K)&(G)&PSNR/SSIM&PSNR/SSIM&PSNR/SSIM&PSNR/SSIM&PSNR/SSIM\\
		\hline
		\multirow{11}{*}{$\times2$}
		&Bicubic&-&-
		&33.64 / 0.9292&30.22 / 0.8683&29.55 / 0.8425&26.87 / 0.8397&30.80 / 0.9339\\
		&SRCNN~\cite{DongC2016TPAMI}&8.0&1.4
		&36.66 / 0.9542&32.47 / 0.9069&31.37 / 0.8879&29.52 / 0.8947&35.73 / 0.9675\\
		&VDSR~\cite{KimJ2016CVPR_VDSR}&664.7&115.1
		&37.53 / 0.9587&33.05 / 0.9127&31.90 / 0.8960&30.77 / 0.9141&37.37 / 0.9737\\
		&LapSRN~\cite{LaiW2019TPAMI}&435.3&18.9
		&37.52 / 0.9591&32.99 / 0.9124&31.80 / 0.8949&30.41 / 0.9101&37.27 / 0.9740\\
		&DRRN~\cite{TaiY2017CVPR}&297.2&1275.5
		&37.74 / 0.9591&33.25 / 0.9137&32.05 / 0.8973&31.23 / 0.9188&37.88 / 0.9750\\
		&IDN~\cite{HuiZ2018CVPR}&715.3&31.0
		&37.83 / 0.9600&33.30 / 0.9148&32.08 / 0.8985&31.27 / 0.9196&38.02 / 0.9749\\
		&CARN~\cite{AhnN2018ECCV}&1592.1&41.9
		&37.90 / 0.9602&33.58 / 0.9175&32.12 / 0.8987&31.95 / 0.9265&38.11 / 0.9764\\
		&IMDN~\cite{HuiZ2019ACMMM}&694.4&29.9
		&{38.00} / \underline{0.9605}&\underline{33.63} / 0.9177&32.19 / 0.8996&{32.13} / 0.9282&38.87 / 0.9774\\
		&CFSRCNN~\cite{TianC2021TMM}&1310.4&87.1
		&37.93 / 0.9603&\underline{33.63} / 0.9176&32.15 / \underline{0.8997}&32.13 / {0.9283}&38.33 / 0.9765\\
		&DeFiAN$_S$~\cite{HuangY2021TIP}&1027.6&44.2
		&\underline{38.03} / \underline{0.9605}&\underline{33.63} / \underline{0.9181}&\underline{32.20} / \underline{0.8997}&\underline{32.20} / \underline{0.9286}&\underline{38.91} / \underline{0.9775}\\
		&U-HDN+IKM(Ours)&1390.9&60.3
		&\bf{38.06} / \bf{0.9607}&\bf{33.77} / \bf{0.9191}&\bf{32.24} / \bf{0.9004}&\bf{32.32} / \bf{0.9301}&\bf{39.09} / \bf{0.9780}\\
		\hline
		\multirow{11}{*}{$\times3$}
		&Bicubic&-&-
		&30.40 / 0.8686&27.54 / 0.7741&27.21 / 0.7389&24.46 / 0.7349&26.95 / 0.8565\\
		&SRCNN~\cite{DongC2016TPAMI}&8.0&1.4
		&32.75 / 0.9090&29.29 / 0.8215&28.41 / 0.7863&26.24 / 0.7991&30.56 / 0.9125\\
		&VDSR~\cite{KimJ2016CVPR_VDSR}&664.7&115.1
		&33.66 / 0.9213&29.78 / 0.8318&28.83 / 0.7976&27.14 / 0.8279&32.13 / 0.9348\\
		&LapSRN~\cite{LaiW2019TPAMI}&435.3&8.5
		&33.82 / 0.9227&29.79 / 0.8320&28.82 / 0.7973&27.07 / 0.8271&32.21 / 0.9344\\
		&DRRN~\cite{TaiY2017CVPR}&297.2&1275.5
		&34.02 / 0.9244&29.98 / 0.8350&28.95 / 0.8004&27.54 / 0.8378&32.72 / 0.9380\\
		&IDN~\cite{HuiZ2018CVPR}&715.3&13.8
		&34.12 / 0.9253&29.99 / 0.8356&28.95 / 0.8013&27.42 / 0.8360&32.71 / 0.9379\\
		&CARN~\cite{AhnN2018ECCV}&1592.1&22.3
		&34.33 / 0.9267&30.30 / 0.8416&29.07 / 0.8044&28.05 / 0.8499&33.32 / 0.9436\\
		&IMDN~\cite{HuiZ2019ACMMM}&703.1&13.5
		&{34.36} / {0.9270}&30.32 / 0.8417&{29.09} / {0.8046}&{28.17} / {0.8519}&{33.61} / {0.9445}\\
		&CFSRCNN~\cite{TianC2021TMM}&1495.0&63.7
		&34.32 / 0.9269&\underline{30.35} / \underline{0.8423}&29.07 / {0.8046}&28.08 / 0.8507&33.44 / 0.9436\\
		&DeFiAN$_S$~\cite{HuangY2021TIP}&1073.7&20.6
		&\underline{34.42} / \bf{0.9273}&{30.34} / {0.8410}&\underline{29.12} / \underline{0.8053}&{\bf 28.20} / \underline{0.8528}&\underline{33.72} / \underline{0.9447}\\
		&U-HDN+IKM(Ours)&1402.4&27.2
		&\bf{34.43} / \bf{0.9273}&\bf{30.39} / \bf{0.8424}&\bf{29.14} / \bf{0.8060}&\underline{28.19} / \bf{0.8537}&\bf{33.85} / \bf{0.9459}\\
		\hline
		\multirow{12}{*}{$\times4$}
		&Bicubic&-&-
		&28.42 / 0.8101&25.99 / 0.7023&25.96 / 0.6672&23.14 / 0.6573&24.89 / 0.7866\\
		&SRCNN~\cite{DongC2016TPAMI}&8.0&1.4
		&30.49 / 0.8629&27.51 / 0.7519&26.91 / 0.7104&24.53 / 0.7230&27.66 / 0.8566\\
		&VDSR~\cite{KimJ2016CVPR_VDSR}&664.7&115.1
		&31.35 / 0.8838&28.02 / 0.7678&27.29 / 0.7252&25.18 / 0.7525&28.87 / 0.8865\\
		&LapSRN~\cite{LaiW2019TPAMI}&435.3&4.8
		&31.52 / 0.8854&28.09 / 0.7687&27.31 / 0.7255&25.21 / 0.7545&29.08 / 0.8883\\
		&DRRN~\cite{TaiY2017CVPR}&297.2&1275.5
		&31.67 / 0.8888&28.22 / 0.7721&27.38 / 0.7284&25.45 / 0.7639&29.44 / 0.8943\\
		&IDN~\cite{HuiZ2018CVPR}&715.3&7.7
		&31.81 / 0.8903&28.25 / 0.7731&27.41 / 0.7295&25.41 / 0.7630&29.42 / 0.8939\\ 
		&CARN~\cite{AhnN2018ECCV}&1592.1&17.1
		&32.15 / \underline{0.8948}&28.58 / 0.7813&27.58 / 0.7360&26.03 / 0.7840&30.34 / {0.9079}\\
		&IMDN~\cite{HuiZ2019ACMMM}&715.2&7.7
		&\underline{32.21} / 0.8946&28.58 / 0.7804&27.56 / 0.7352&26.04 / 0.7838&{30.45} / 0.9072\\
		&MS$^3$-Conv~\cite{FengR2020ECCV}
		&1300.0&21.6
		&32.09 / 0.8945&28.61 / \underline{0.7821}&\underline{27.59} / \underline{0.7368}&{26.09} / {0.7859}&- / -\\
		&CFSRCNN~\cite{TianC2021TMM}&1458.1&58.7
		&32.10 / 0.8932&\underline{28.63} / 0.7814&27.55 / 0.7346&26.06 / 0.7836&30.43 / 0.9068\\
		&DeFiAN$_S$~\cite{HuangY2021TIP}&1064.5&12.8
		&{32.16} / {0.8942}&\underline{28.63} / {0.7810}&{27.58} / {0.7363}&\underline{26.10} / \underline{0.7862}&\underline{30.59} / \underline{0.9084}\\
		&U-HDN+IKM(Ours)&1400.1&15.7
		&\bf{32.29} / \bf{0.8955}&\bf{28.68} / \bf{0.7826}&\bf{27.62} / \bf{0.7373}&\bf{26.17} / \bf{0.7880}&\bf{30.85} / \bf{0.9110}\\
		\hline
		\hline
	\end{tabular}
	\label{tab:full-reference_comparison}
\end{table*}

\section{Experiments}
\subsection{Datasets and Evaluation}
Following the existing single image SR researches, we implement our experiments on several benchmark datasets for evaluation: Set5~\cite{BevilacC2012BMVC}, Set14~\cite{ZeydeR2010}, BSD100~\cite{ArbelaezP2011TPAMI} and Manga109~\cite{LaiW2019TPAMI} datasets.
And in the training phase, 3450 high-quality images from DIV2K and Flickr2K~\cite{TimofteR2017CVPRW} dataset are considered and down-sampled using Bicubic algorithm to generate the HR-LR pairs. In detail, we use the $48\times48$ RGB patches from the low-resolution training set as input and the corresponding $48s\times48s$ high-resolution patches as ground-truth for $\times s$ upscaling, and augment these LR-HR pairs with randomly horizontal flips and 90 rotations. Particularly, all the LR and HR images are pre-processed by subtracting the mean RGB value of the training sets.

To evaluate the SR performance, we apply two full-reference image quality assessment (IQA) criteria: Peak Signal-to-Noise Ratio (PSNR) and Structural SIMilarity (SSIM). And following the convention of SR, only the luminance channel is selected for full-reference image quality assessment because the intensity of image is more sensitive to human vision than the chroma channels. Moreover, we use two criteria to represent the computational complexities: \#Params (space complexity) denotes the number of parameters, and \#FLOPs (time complexity) indicates the number of operations by Multi-Adds which is the number of composite multiply-accumulate operations for generating a $480\times360\times3$ output.

\subsection{Implementation Details}
\subsubsection{Hyperparameters}~\label{sec:Hyperparameters}
As illustrated in Section~\ref{sec:UHDN}, we stack $N$=$4$ U-HDBs to build a lightweight U-HDN with only 1.4M parameters. And in each U-HDB, by applying U-style residual learning and hourglass dense block learning, $M$=$6$ dense units are stacked with symmetric depth of $[6,5,4,4,5,6]$, where the growth rate $g$=$12$ and input channel in head layer $C$=$64$. Furthermore, IKM is applied into each layer of U-HDN except for the transition layers, and the threshold of $\boldsymbol{f}_{PIA}$ is set to $t$=$0$. Besides, we utilize the sub-pixel convolutions~\cite{ShiW2016CVPR} for feature upscaling in the upscaling module as in Fig.~\ref{fig:architecture of UHDN}.

\subsubsection{Optimization}
All of our models are optimized via minimizing the mean absolute error (MAE) between the super-resolved image $\boldsymbol{f}_{model}(\boldsymbol{x})$ and the corresponding ground-truth HR $\boldsymbol{y}$. Therefore, given a training dataset $\{\boldsymbol{x}^b, \boldsymbol{y}^b\}^B_{b=1}$, where $B$ is the batch size and $\{\boldsymbol{x}^b, \boldsymbol{y}^b\}$ are the $b$-th LR and HR patch pairs. Then, the objective function is formulated as
\begin{equation}
\boldsymbol{\mathcal{L}}(\Theta)=\sum_{b=1}^{B}\left \| \boldsymbol{y}^b - \boldsymbol{f}_{model}(\boldsymbol{x}^b; \Theta)  \right \|_1
\label{eq:objective function}
\end{equation}
where $\Theta$ denotes the trainable parameters of models, which are optimized by using the Adam optimizer~\cite{KingmaD2014ICLR} with mini-batches of size $B=16$, with the learning rate being initialized to ${{10}^{-4}}$ and halved for every $10^5$ mini-batch updates. Each of the final models will get convergence after $3\times10^5$ updates on the PyTorch framework and a 32GB NVIDIA Tesla V100 GPU.

\subsection{Comparison with State-of-the-art Methods}
\begin{figure*}[!t]
	\centering
	\vspace{-0.2cm}
	\begin{minipage}[b]{1\linewidth}
		\centering
		\subfloat[HR Ground-truth\protect\\PSNR / SSIM]{
			\includegraphics[width=0.3\linewidth]{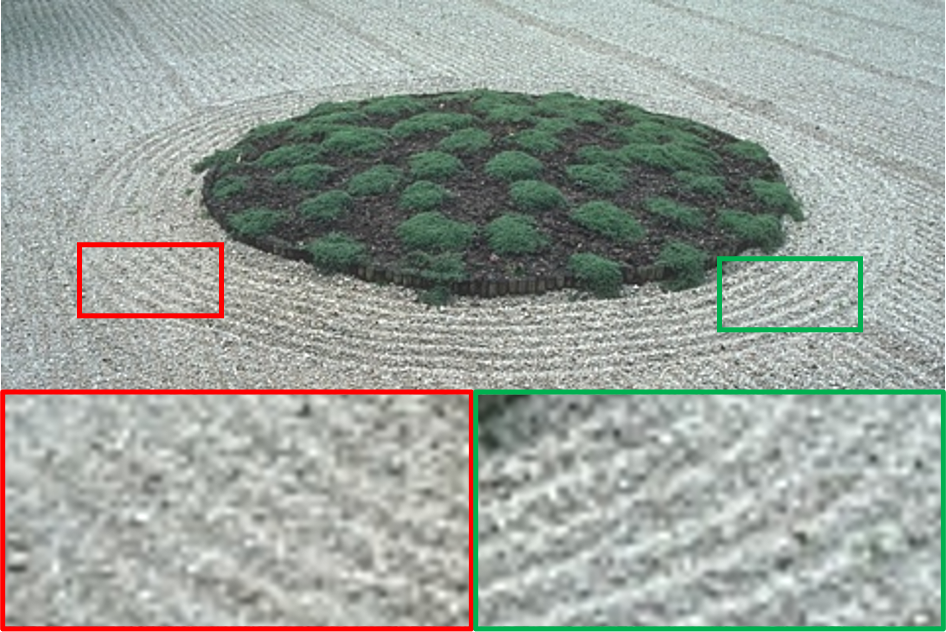}
		}
		\subfloat[Bicubic\protect\\ 21.78 / 0.2696]{
			\includegraphics[width=0.3\linewidth]{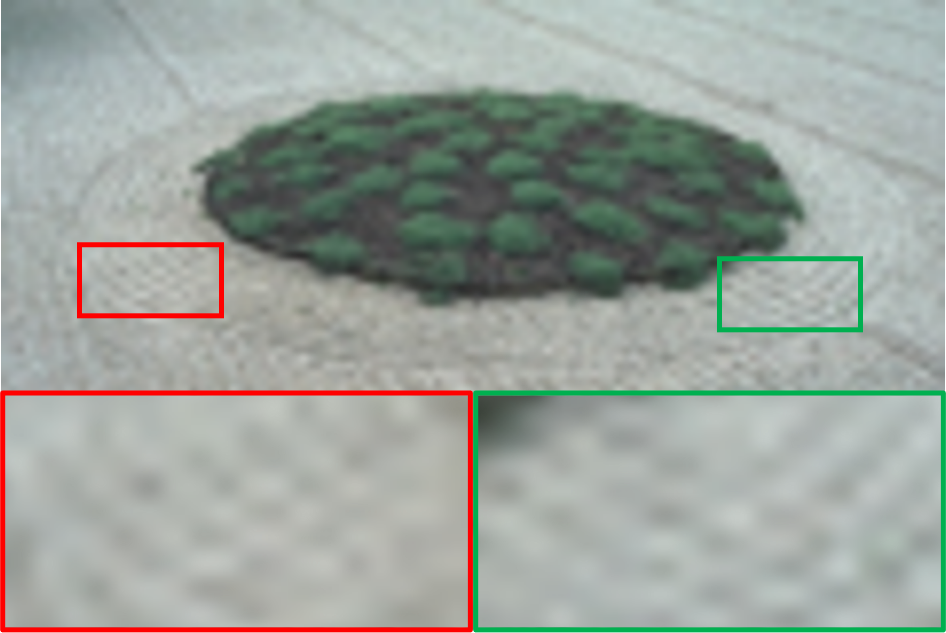}
		}
		\subfloat[SRCNN~\cite{DongC2016TPAMI}\protect\\ 21.87 / 0.3051]{
		\includegraphics[width=0.3\linewidth]{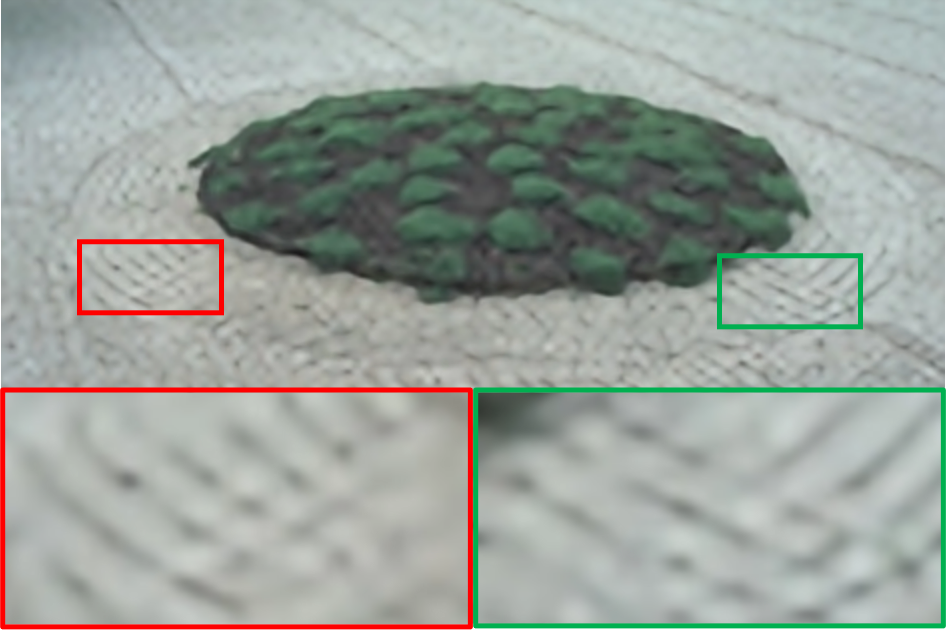}
		}\\
		\vspace{-0.3cm}
		\subfloat[VDSR~\cite{KimJ2016CVPR_VDSR}\protect\\ 21.90 / 0.3115]{
			\includegraphics[width=0.3\linewidth]{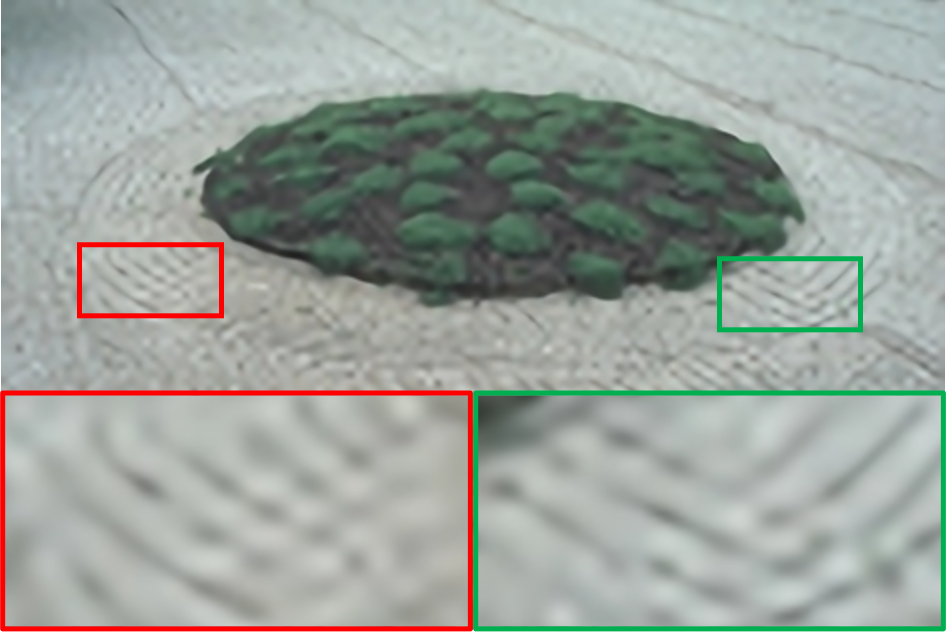}
		}
		\subfloat[CARN~\cite{AhnN2018ECCV}\protect\\ 21.90 / 0.3136]{
		\includegraphics[width=0.3\linewidth]{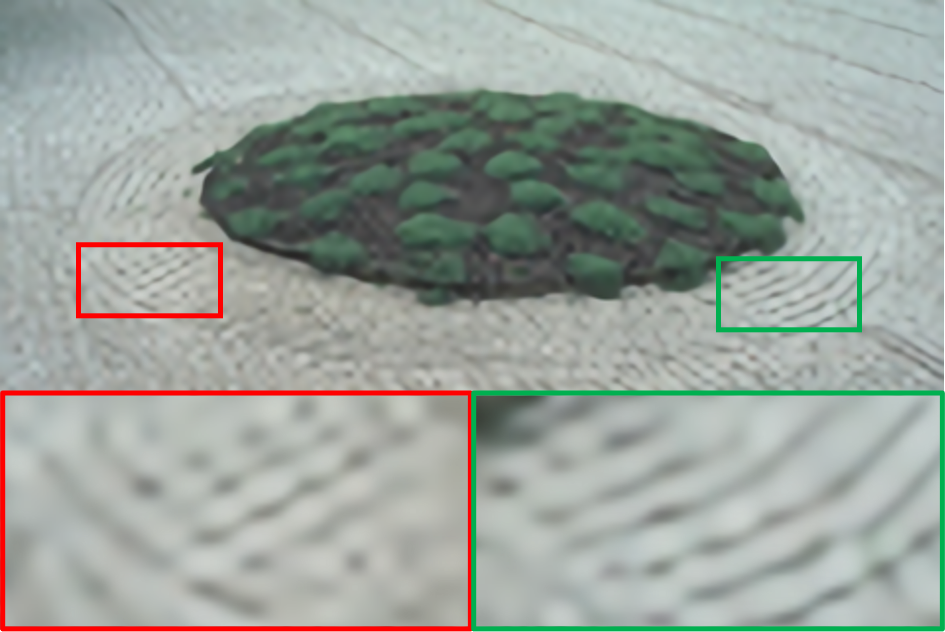}
		}
		\subfloat[U-HDN+IKM (Ours)\protect\\ \textbf{21.94} / \textbf{0.3176}]{
		\includegraphics[width=0.3\linewidth]{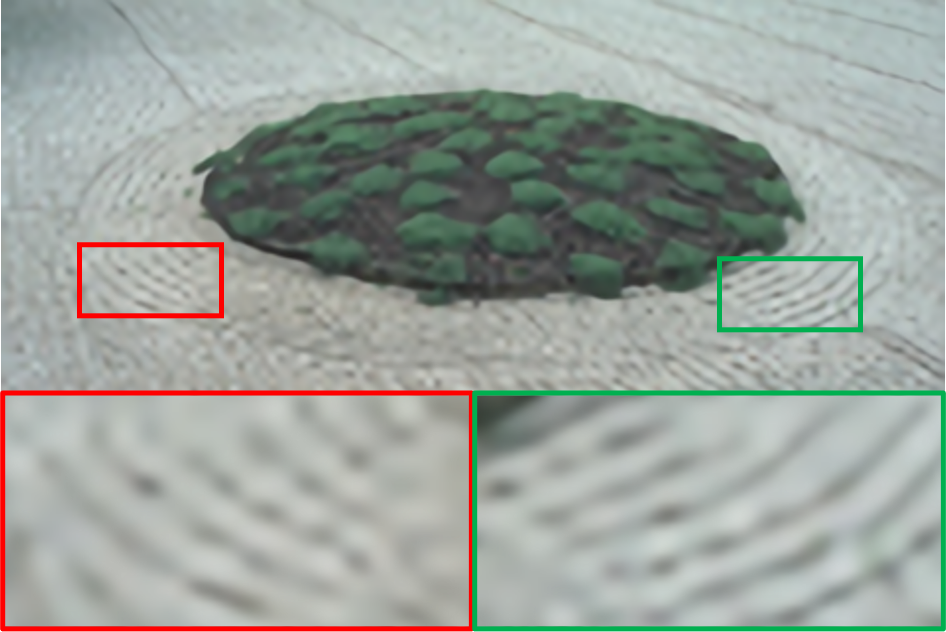}
		}
	\end{minipage}
	\caption{Subjective quality assessment for $\times4$ upscaling on image: ``{\em 86016}" from BSD100 dataset. Obviously, the global composition of this image is lying on the center region with lower illuminance, so the IKM attention weight seems as an ``isotropic Laplace operator'' and keeps the edges well-preserved.}
	\label{fig:Subjective comparisons: 1}
\end{figure*}

\begin{figure*}[!h]
	\centering
	\begin{minipage}[b]{0.81\linewidth}
		\centering
		\vspace{-0.5cm}
		\subfloat[HR Ground-truth\protect\\PSNR / SSIM]{
			\includegraphics[width=0.3\linewidth]{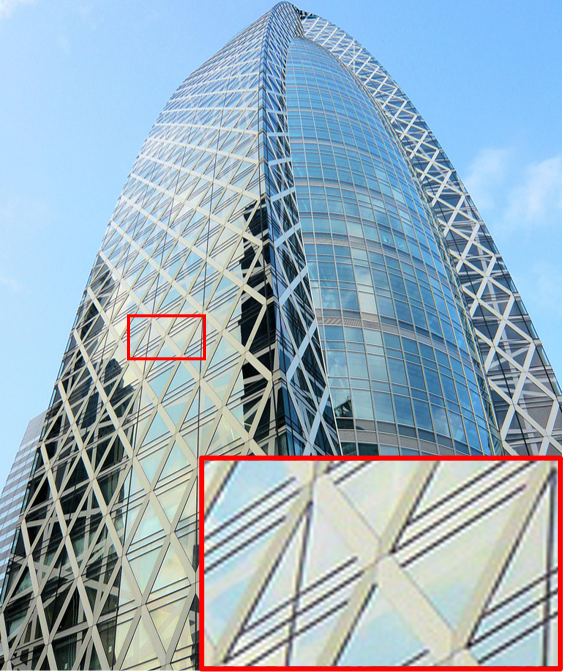}
		}
		\subfloat[Bicubic\protect\\20.86 / 0.5794]{
			\includegraphics[width=0.3\linewidth]{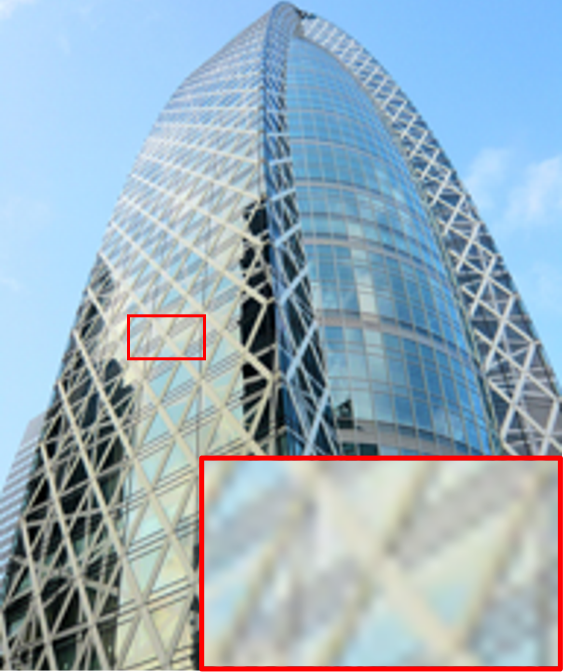}
		}
		\subfloat[SRCNN~\cite{DongC2016TPAMI}\protect\\ 21.69 / 0.6464]{
			\includegraphics[width=0.3\linewidth]{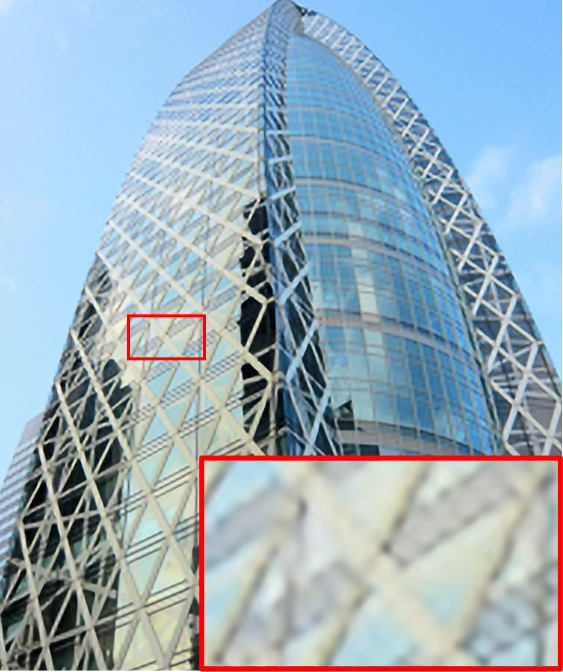}
		}\\
		\vspace{-0.3cm}
		\subfloat[VDSR~\cite{KimJ2016CVPR_VDSR}\protect\\ 22.26 / 0.6874]{
			\includegraphics[width=0.3\linewidth]{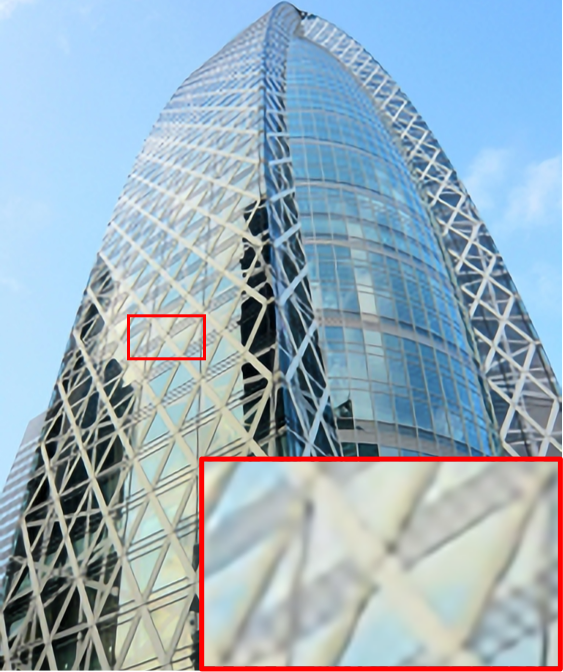}
		}
		\subfloat[CARN~\cite{AhnN2018ECCV}\protect\\ 23.18 / 0.7482]{
			\includegraphics[width=0.3\linewidth]{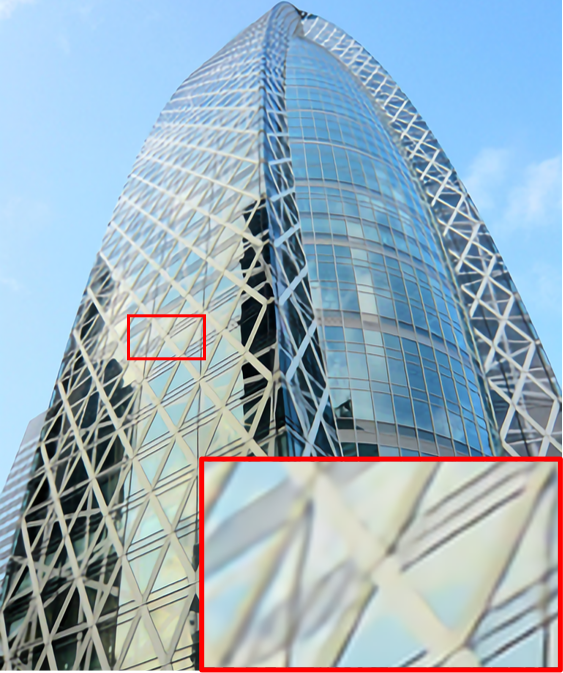}
		}
		\subfloat[U-HDN+IKM(Ours)\protect\\ \textbf{23.23} / \textbf{0.7535}]{
			\includegraphics[width=0.3\linewidth]{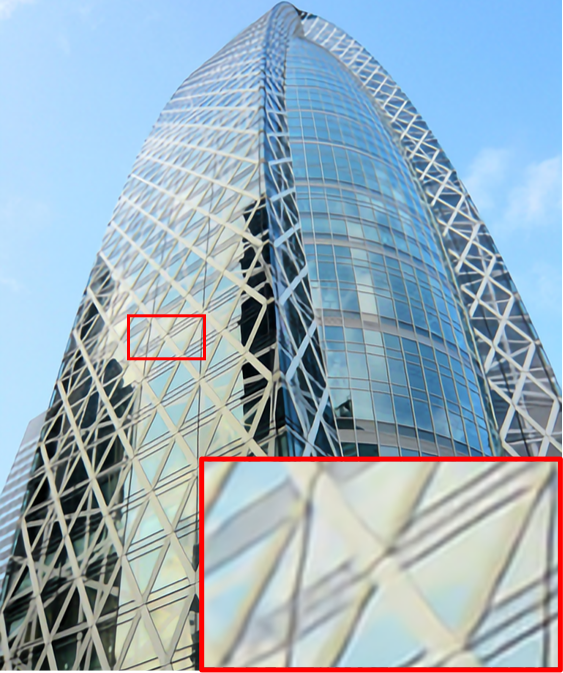}
		}
	\end{minipage}
	\caption{Subjective quality assessment for $\times4$ upscaling on image: ``{\em img039}" from Urban100 dataset. Obviously, the global composition of this image shows higher contrasts in the vertical direction, so the IKM attention weight seems as an ``horizontal Prewitt operator'', which suppresses the distortions and keeps the vertical edges well-preserved.}
	\label{fig:Subjective comparisons: 2}
\end{figure*}
\begin{table*}
	\renewcommand\arraystretch{1.1}
	\centering
	\captionsetup{justification=centering}
	\caption{\textsc{\\Quantitative comparisons with the state-of-the-art high-fidelity SISR methods for $\times2$ upscaling.}}
		\begin{tabular}{l c c c c c c c}
			\hline
			\hline
			\multirow{2}{*}{Method}&\#Params&\#FLOPs
			&{Set5}&{Set14}&{BSD100}&{Urban100}&{Manga109}\\
			&(M)&(G)&PSNR/SSIM&PSNR/SSIM&PSNR/SSIM&PSNR/SSIM&PSNR/SSIM\\
			\hline
			Bicubic&-&-
			&33.64 / 0.9292&30.22 / 0.8683&29.55 / 0.8425&26.87 / 0.8397&30.80 / 0.9339\\
			U-HDN+IKM(Ours)&1.39&60.3
			&{38.06} / {0.9607}&{33.77} / {0.9191}&{32.24} / {0.9004}&{32.32} / {0.9301}&{39.09} / {0.9780}\\
			\hline
			EDSR~\cite{LimB2017CVPRW}&40.7&1760.0
			&38.11 / 0.9600&33.82 / 0.9189&32.33 / 0.9011&32.94 / 0.9351&39.10 / 0.9773\\
			RDN~\cite{ZhangY2018CVPR}&5.62&243.0
			&38.16 / 0.9603&33.88 / 0.9199&32.31 / 0.9009&32.89 / 0.9353&39.09 / 0.9771\\
			RCAN~\cite{ZhangY2018ECCV}&15.4&663.5
			&\underline{38.18} / 0.9604&\underline{34.00} / \underline{0.9203}&\textbf{32.37} / \underline{0.9016}&
			\bf{33.14} / \bf{0.9364}&\underline{39.34} / \underline{0.9777}\\
			OISR-RK3~\cite{HeX2019CVPR}&44.3&1812.1
			&38.13 / 0.9602&33.81 / 0.9194&32.34 / 0.9011&33.00 / 0.9357&39.13 / 0.9773\\
			CSFM~\cite{HuY2020TCSVT}&12.1&519.9
			&38.17 / \underline{0.9605}&33.94 / 0.9200&32.34 / 0.9013&\underline{33.08} / \underline{0.9358}&39.30 / 0.9775\\
			U-HDN$_L$+IKM(Ours)&10.9&471.5
			&\textbf{38.23} / \textbf{0.9614}&\textbf{34.03} / \textbf{0.9217}&\underline{32.35} / \textbf{0.9018}&32.84 / 0.9346&\textbf{39.46} / \textbf{0.9788}\\
			\hline
			\hline
	\end{tabular}
	\label{tab:full-reference_comparison_large}
	\vspace{-0.1cm}
\end{table*}

\subsubsection{Comparison of full models}
As described in Section~\ref{sec:Hyperparameters}, the default settings are defined for lightweight applications, then several state-of-the-art lightweight (\#Params$\le$2.0M or \#FLOPs$\le$100G) SR methods (\eg, SRCNN~\cite{DongC2016TPAMI}, VDSR~\cite{KimJ2016CVPR_VDSR}, LapSRN~\cite{LaiW2019TPAMI}, DRRN~\cite{TaiY2017CVPR}, IDN~\cite{HuiZ2018CVPR}, CARN~\cite{AhnN2018ECCV}, IMDN~\cite{HuiZ2019ACMMM}, MS$^3$-Conv~\cite{FengR2020ECCV}, CFSRCNN~\cite{TianC2021TMM} and DeFiAN$_S$~\cite{HuangY2021TIP}) are selected to compare with our ``UHDN+IKM''. Besides, based on the default settings in Section~\ref{sec:Hyperparameters}, we also build a larger model ``UHDN$_L$+IKM'' by resetting $N$=$8$, $g$=$24$ and $C$=$128$ for high-fidelity application, and compare it with the state-of-the-art high-fidelity SR methods (\eg, EDSR~\cite{LimB2017CVPRW}, RDN~\cite{ZhangY2018CVPR}, RCAN~\cite{ZhangY2018ECCV}, OISR-RK3~\cite{HeX2019CVPR} and CSFM~\cite{HuY2020TCSVT}).

To evaluate the pixel-wise information fidelity and the computational complexities, we conduct the experimental comparisons on the proposed method and other state-of-the-art SISR methods, and the quantitative full-reference IQA results are reported in TABLE~\ref{tab:full-reference_comparison} and TABLE~\ref{tab:full-reference_comparison_large}. Particularly, since no additional parameters need to be optimized in IKM, both of the lightweight ``U-HDN+IKM'' and high-fidelity ``U-HDN$_L$+IKM'' achieve relative best performances with lower computational complexities against other state-of-the-art SISR methods.
Moreover, we also post up the super-resolved results in chromatic format for subjective quality assessment as Fig.~\ref{fig:Subjective comparisons: 1} and Fig.~\ref{fig:Subjective comparisons: 2} show. By introducing the image-specific attentions to modulate the convolutional kernels, our methods are adaptive to reconstruct different images with the corresponding specific kernels. Particularly, in Fig.~\ref{fig:Subjective comparisons: 1}, the global contextual composition of the image is lying on the center with lower illuminance, so the IKM attention weight acts as an ``isotropic Laplace operator'' and keeps the edges well-preserved. And in Fig.~\ref{fig:Subjective comparisons: 2}, the composition of the image shows higher contrasts in vertical directions, so the IKM attention weight acts as an ``horizontal Prewitt operator'', which suppresses the distortions and keeps the vertical edges well-preserved.

\begin{figure*}
	\begin{center}
		\includegraphics[width=0.73\linewidth]{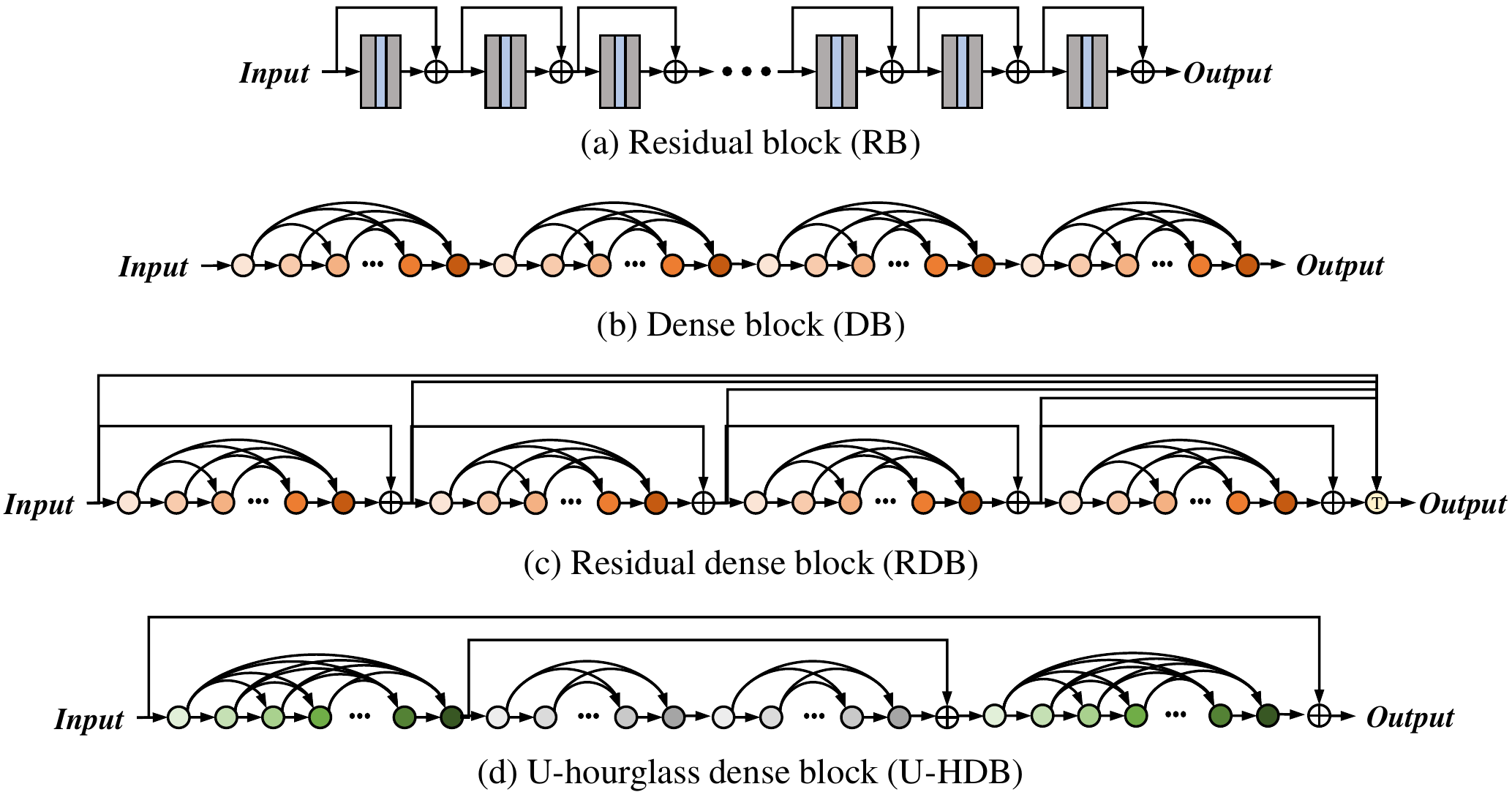}
	\end{center}
	\caption{Comparison of U-hourglass dense block (U-HDB) and other state-of-the-art architectures, including: (a) residual block (RB) in EDSR~\cite{LimB2017CVPRW}; (b) dense block (DB) in SRDenseNet~\cite{TongT2017ICCV} and MemNet~\cite{TaiY2017ICCV}; (c) residual dense block (RDB) in RDN~\cite{ZhangY2018CVPR}.}
	\label{fig:UHDB}
\end{figure*}
\subsubsection{Comparison of architectures}
Aiming at demonstrating the effectiveness of the proposed architectures (U-HDN and IKM), we apply several state-of-the-art architectures into the same backbone by stacking $N=1$ blocks, and then train these models with $5\times10^4$ mini-batch updates. 
The architectures are shown in Fig.~\ref{fig:UHDB} and specified as following:

\begin{table}
	\renewcommand\arraystretch{1.15}
	\centering
	\captionsetup{justification=centering}
	\caption{\textsc{\\Comparisons with state-of-the-art \\architectures for $\times2$ upscaling on Set5 dataset.}}
	\resizebox{1\linewidth}{!}{
		\begin{tabular}{{p{0.9cm} p{1.4cm}|p{0.5cm} p{0.4cm}|p{0.3cm} p{0.3cm}|c}}
			\hline
			\hline
			\multicolumn{2}{c|}{Architectures}&\multicolumn{2}{c|}{\#Params (K)}&\multicolumn{2}{c|}{\#FLOPs (G)}&{PSNR} / {SSIM}\\
			\hline
			\multirow{4}{*}{RB\cite{LimB2017CVPRW}}&-&\multirow{4}{*}{362.1}&+0&\multirow{4}{*}{15.8}&+0&37.47 / 0.9584\\
			&+CA\cite{ZhangY2018ECCV}&&+19.4&&+$\approx$0&37.53 / 0.9586\\
			&+SA\cite{HuY2020TCSVT}&&+9.62&&+0.42&37.52 / 0.9586\\
			&+IKM~(Ours)&&+0&&+$\approx$0&37.52 / 0.9586\\
			\hline
			\multirow{4}{*}{DB\cite{TongT2017ICCV}}&-&\multirow{4}{*}{382.3}&+0&\multirow{4}{*}{16.7}&+0&37.44 / 0.9583\\
			&+CA\cite{ZhangY2018ECCV}&&+2.61&&+$\approx$0&37.46 / 0.9585\\
			&+SA\cite{HuY2020TCSVT}&&+1.29&&+0.06&37.48 / 0.9585\\
			&+IKM~(Ours)&&+0&&+$\approx$0&37.52 / 0.9586\\
			\hline
			\multirow{4}{*}{RDB\cite{ZhangY2018CVPR}}&-&\multirow{4}{*}{411.0}&+0&\multirow{4}{*}{18.0}&+0&37.48 / 0.9584\\
			&+CA\cite{ZhangY2018ECCV}&&+2.61&&+$\approx$0&37.48 / 0.9586\\
			&+SA\cite{HuY2020TCSVT}&&+1.29&&+0.06&37.53 / 0.9586\\
			&+IKM~(Ours)&&+0&&+$\approx$0&37.55 / 0.9587\\
			\hline
			\multirow{4}{*}{\tabincell{c}{U-HDB\\(Ours)}}&-&\multirow{4}{*}{384.9}&+0&\multirow{4}{*}{16.9}&+0&37.48 / 0.9585\\
			&+CA\cite{ZhangY2018ECCV}&&+2.61&&+$\approx$0&37.47 / 0.9585\\
			&+SA\cite{HuY2020TCSVT}&&+1.29&&+0.06&37.52 / 0.9586\\
			&+IKM~(Ours)&&+0&&+$\approx$0&\textbf{37.57} / \textbf{0.9588}\\
			\hline
			\hline
		\end{tabular}
	}
	\label{tab:comparison architectures refconv}
\end{table}
{\em i) Residual block (RB)}. To train very deep networks, residual learning is introduced for image classification in ResNet~\cite{HeK2016CVPR} and extended for image super-resolution (SR) in EDSR~\cite{LimB2017CVPRW} and DRRN~\cite{TaiY2017CVPR}. In this section, we stack $M=8$ enhanced residual block (RB) with $C=48$ in each residual block to build a deep but relatively lightweight network to compare with our U-HDB.

{\em ii) Dense block (DB)}. To improve the feature diversity, the dense connection is proposed for visual classification in DenseNet~\cite{HuangG2017CVPR} and extended for SR in SRDenseNet~\cite{TongT2017ICCV}. Similar to SRDenseNet but with some modifications, we apply a group of dense units with the same depth $d$=5 in each block of backbone, and set the growth rate $g$=12 as our U-HDN.

{\em iii) Residual dense block (RDB)}. To exploit both characteristics of residual learning and dense connection, Zhang~\etal\cite{ZhangY2018CVPR} propose the residual dense network with RDBs, which achieves the state-of-the-art performance. Different from SRDenseNet, RDN embeds the dense unit into the residual learning structure, and then utilizes the global feature fusion module (implemented as the transition layer $\boldsymbol{f}_{Transition}(\cdot)$ in DenseNet) in the tail of network for intermediate features fusion. For a fair comparison, we apply the backbone of RDN in each block and set the growth gate $g$=12.

{\em iv) U-hourglass dense block (U-HDB)}. Aiming at further improving the feature diversity, we propose the U-HDB by stacking symmetrically hourglass dense units with U-style residual learning, which is expected to be an appropriate architecture for utmost improvement of utilizing IKM theoretically and achieves the state-of-the-art performance against the prior methods as illustrated in the paper.

As in TABLE~\ref{tab:comparison architectures refconv}, even using the vanilla conventional convolution, the proposed U-HDB reaches relatively better performance than the other state-of-the-art architectures, and by further applying the IKM, all the architectures achieve relatively higher improvements than applying channel attention (CA) or spatial attention (SA).

\subsection{Model Analysis}
In this subsection, we mainly conduct some ablation studies and model analysis with experimental demonstrations.
Specifically, the models in this section are trained with $5\times10^4$ mini-batch updates on DIV2K datasets under the settings as $N=4$, $M=6$ with depth of $[6,5,4,4,5,6]$, $C=64$, $g=12$, unless otherwise specified.

\subsubsection{Effect of model size}
\begin{figure}
	\includegraphics[width=1\linewidth]{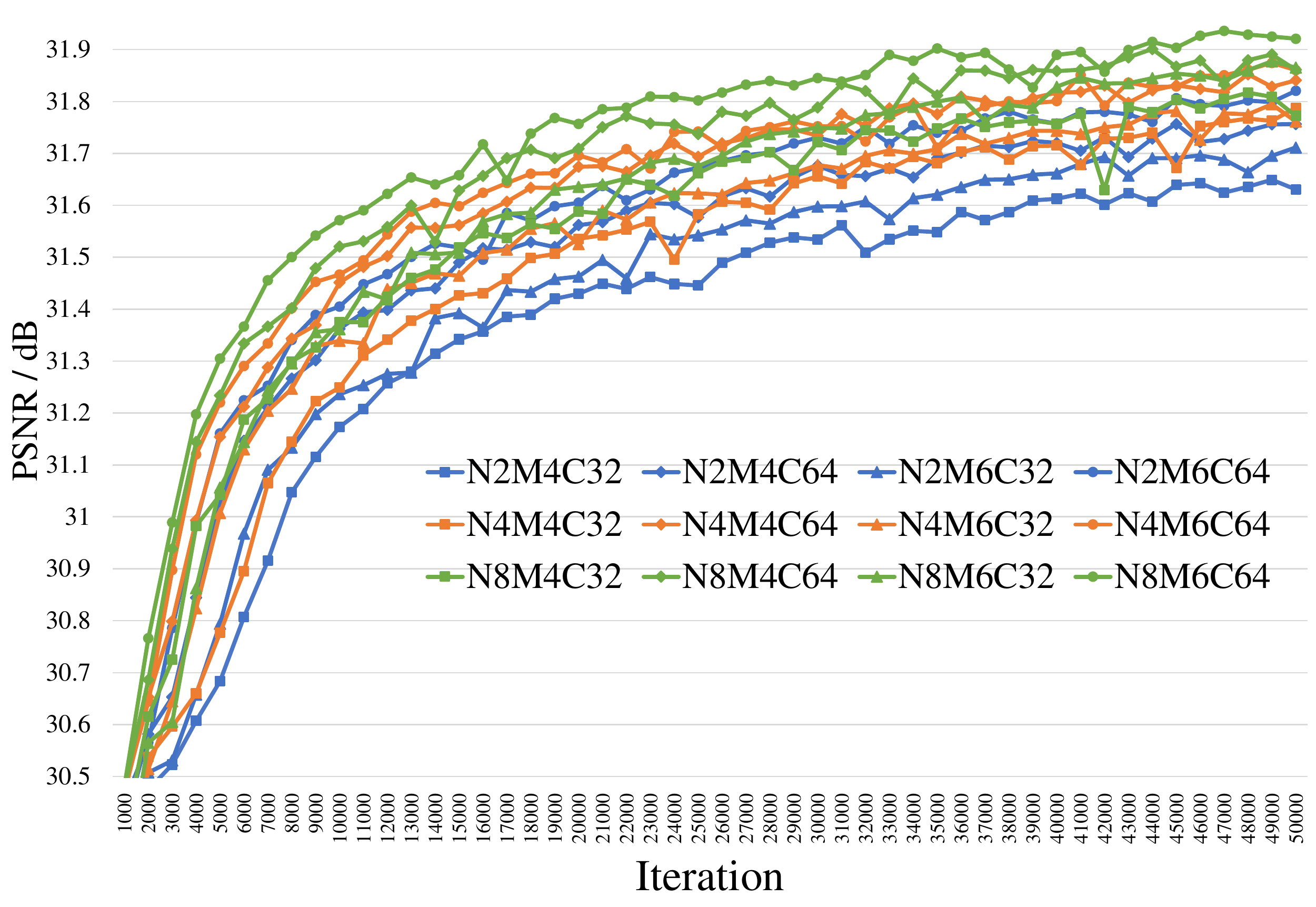}
	\caption{Effect of model size with different hyperparameters: average PSNR on BSD100 dataset for $\times2$ upscaling. Note that, ``N4M6C64'' represents the U-HDN (w/ IKM) with $N=4$ U-HDBs, in each U-HDB, $M=6$ dense units are stacked with depthes of $[6,5,4,4,5,6]$ and the number of channels are $C=64$.}
	\label{fig:effect_model_size}
\end{figure}

As illustrated in Section~\ref{sec:Hyperparameters}, our U-HDN (w/ IKM) is designed under several hyperparameters, including the number of U-HDBs $N$, the number of dense units in each U-HDB $M$, and the input channel in the head of dense unit $C$. We then conduct the model analysis on different settings of hyperparameters as in Fig.~\ref{fig:effect_model_size}. In detail, the symmetric depths of U-HDBs are set to $[6,5,4,4,5,6]$ when $M=6$ or $[6,4,4,6]$ when $M=4$.

As shown in Fig.~\ref{fig:effect_model_size}, we find that, as the model size increases (\ie, larger $N$, $M$, $C$ and symmetric depth of each U-HDB), the performance gets a large margin of improvement but reaches the bottleneck when the model size reaches a peak level. Therefore, we choose an appropriate setting of hyperparameter as described in Section~\ref{sec:Hyperparameters}, which achieves excellent performance with relatively lower computational complexities.

\subsubsection{Effect of IKM}
As illustrated in Section~\ref{sec:IKM}, IKM is regarded as enhancing the input features with the modulated kernels as in Eqs.~(\ref{eq:reflective convolution}-\ref{eq:attention generation}), we then conduct several investigations on the effects of the whole IKM module, the attention generation module $\boldsymbol{f}_{CAG}$, and the size of receptive field $\mathcal{R}$.

{\em i) Investigations on performance}.
To demonstrate the effect of the proposed IKM, we apply the vanilla convolution, the variants with channel attention (CA), spatial attention (SA), or IKM into several architectures in Fig.~\ref{fig:UHDB}. 
As reported in TABLE~\ref{tab:comparison architectures refconv}, by applying our IKM, although without any additional parameters, the models possess a large margin of improvement for all the implemented architectures over the vanilla convolution, and achieve comparable performance with relatively lower complexity against the variants with channel attention and spatial attention.
Particularly, as aforementioned in Section~\ref{sec:UHDN}, dense connection should be expected to be more appropriate for utmost improvement of utilizing IKM. We find the larger margin of improvements are arisen on the dense connections, \ie, 0.05dB for RB, but 0.08dB for DB, 0.07dB for RDB, and 0.09dB for U-HDB.

\begin{figure}
	\begin{center}
		\includegraphics[width=0.95\linewidth]{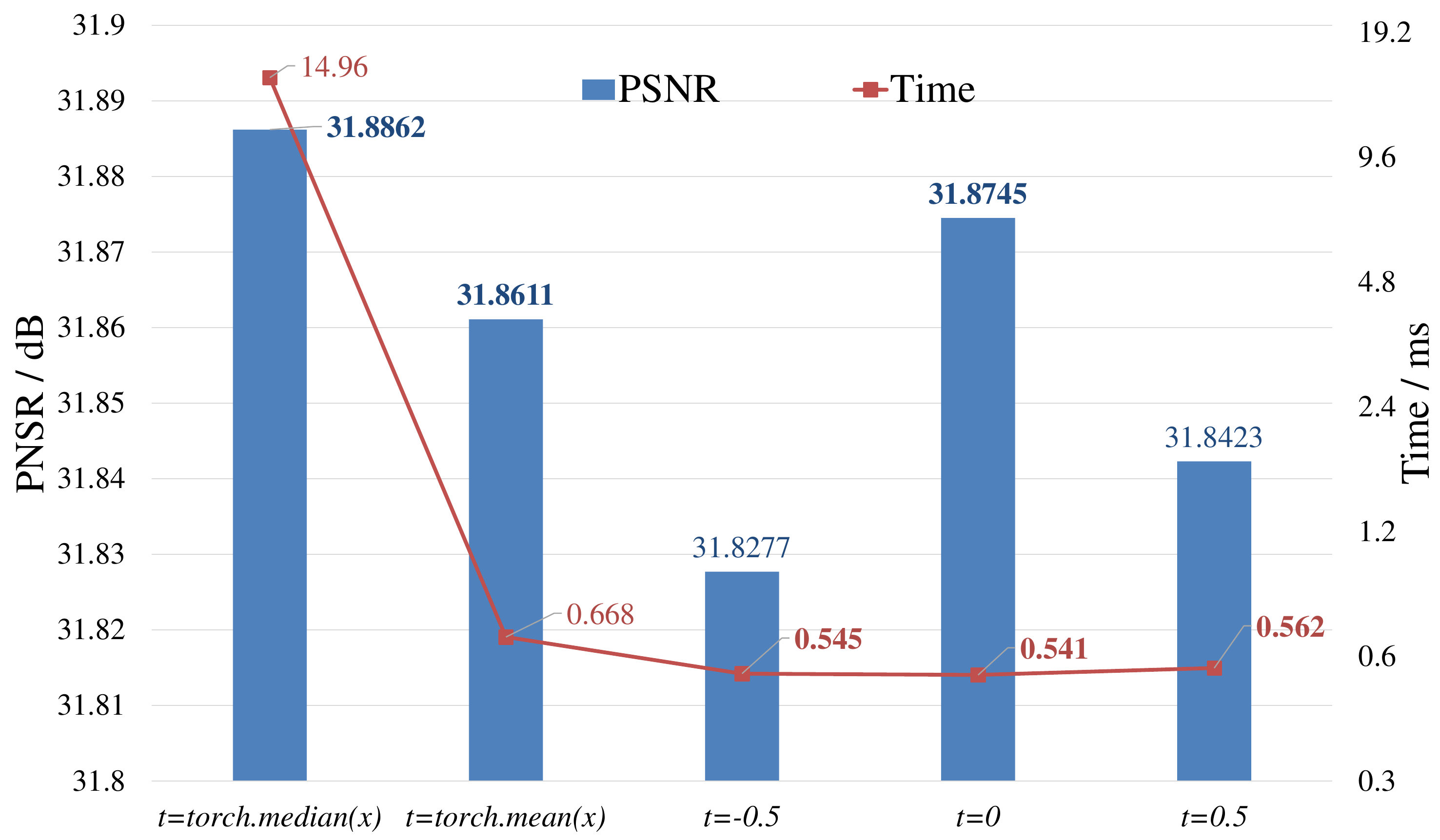}
	\end{center}
	\caption{Investigation on threshold $t$ of $\boldsymbol{f}_{PIA}$: average PSNR and execution time of $\boldsymbol{f}_{PIA}$ on BSD100 dataset for $\times2$ upscaling.}
	\label{fig:investigation_threshold}
\end{figure}

{\em ii) Investigation on $\boldsymbol{f}_{CAG}$}. To generate the image-specific layout of features, we introduce a layout generation module $\boldsymbol{f}_{CAG}$ into each IKM. So we also conduct investigations on the setting of $\boldsymbol{f}_{CAG}$, specifically, on the threshold in $\boldsymbol{f}_{PIA}$.
As in Eq.~(\ref{eq:PIA}) and Fig.~\ref{fig:architecture of CAG}, a hard threshold function is applied to sample the principal information of feature. Since the whole U-HDN model and U-HDBs work in a residual learning manner as in Fig.~\ref{fig:architecture of UHDN} and Fig.~\ref{fig:UHDB}, most values in residual features are likely to be zero as described in~\cite{KimJ2016CVPR_VDSR}. So we suggest $t=0$ as the threshold to activate the principal information and next to sample them. And to demonstrate this assumption, we conduct some investigations on choosing an appropriate threshold as in Fig.~\ref{fig:investigation_threshold}. Specifically and intuitively, the median value of feature should be more appropriate than 0, however, median operation $torch.median()$ need to sort all atoms in each location of feature map, and is inefficient in parallel computing with GPUs especially for large patches.

\begin{table}
	\renewcommand\arraystretch{1.05}
	\centering
	\captionsetup{justification=centering}
	\caption{\textsc{\\Investigation on local receptive field $\mathcal{R}$\\ for $\times2$ upscaling on BSD100 dataset.}}
	\resizebox{1\linewidth}{!}{
		\begin{tabular}{c c c|c c c c c}
			\hline
			\hline
			{Kernel}&{Dilation}&\multirow{2}{*}{$\mathcal{R}$}&\multirow{2}{*}{IKM}&\multirow{2}{*}{\#Params}&\multirow{2}{*}{PSNR}&\multirow{2}{*}{SSIM}\\
			(K)&(D)&&&&&\\
			\hline
			\multirow{2}{*}{$3\times3$}&\multirow{2}{*}{$1\times1$}&\multirow{2}{*}{$3\times3$}&$\times$&1390.9&31.80&0.8951\\
			&&&{\checkmark}&{1390.9}&31.87&0.8960\\
			\hline
			\multirow{2}{*}{$3\times3$}&\multirow{2}{*}{$2\times2$}&\multirow{2}{*}{$5\times5$}&$\times$&1390.0&31.72&0.8945\\
			&&&{\checkmark}&{1390.0}&31.79&0.8954\\
			\hline
			\multirow{2}{*}{$3\times3$}&\multirow{2}{*}{$3\times3$}&\multirow{2}{*}{$7\times7$}&$\times$&1390.0&31.61&0.8927\\
			&&&{\checkmark}&{1390.9}&31.71&0.8942\\
			\hline
			\multirow{2}{*}{$5\times5$}&\multirow{2}{*}{$1\times1$}&\multirow{2}{*}{$5\times5$}&$\times$&3436.7&31.82&0.8952\\
			&&&{\checkmark}&{3436.7}&31.91&0.8968\\
			\hline
			\multirow{2}{*}{$5\times5$}&\multirow{2}{*}{$2\times2$}&\multirow{2}{*}{$9\times9$}&$\times$&3436.7&31.69&0.8936\\
			&&&{\checkmark}&{3436.7}&31.83&0.8954\\
			\hline
			\hline
			\multicolumn{7}{l}{*$\mathcal{R}=K+(K-1)*(D-1)$.}
	\end{tabular}}
	\vspace{-0.4cm}
	\label{tab:investigation_recep_field}
\end{table}
{\em iii) Investigation on local receptive field $\mathcal{R}$.}
As illustrated in Section~\ref{sec:IKM} and Eq.~(\ref{eq:reflective convolution}), IKM is equivalent to introducing attentions to act on the local receptive fields of feature, we then investigate several settings of local receptive field while with or without IKM. In addition, in deep full convolutional networks, the receptive field depends on the settings of convolutional kernel (\eg, ``kernel size'', ``dilation'') and is calculated as $\mathcal{R}=K+(K-1)*(D-1)$~\cite{YuF2016ICLR}. As reported in TABLE~\ref{tab:investigation_recep_field}, with the local receptive field increases, IKM achieves higher gains on both PSNR and SSIM, \eg, when $\mathcal{R}=9\times9$, the gain reaches 0.14dB on PSNR.


\begin{figure*}
	\centering
	\subfloat[B=2]{
		\footnotesize
		\includegraphics[width=0.45\linewidth]{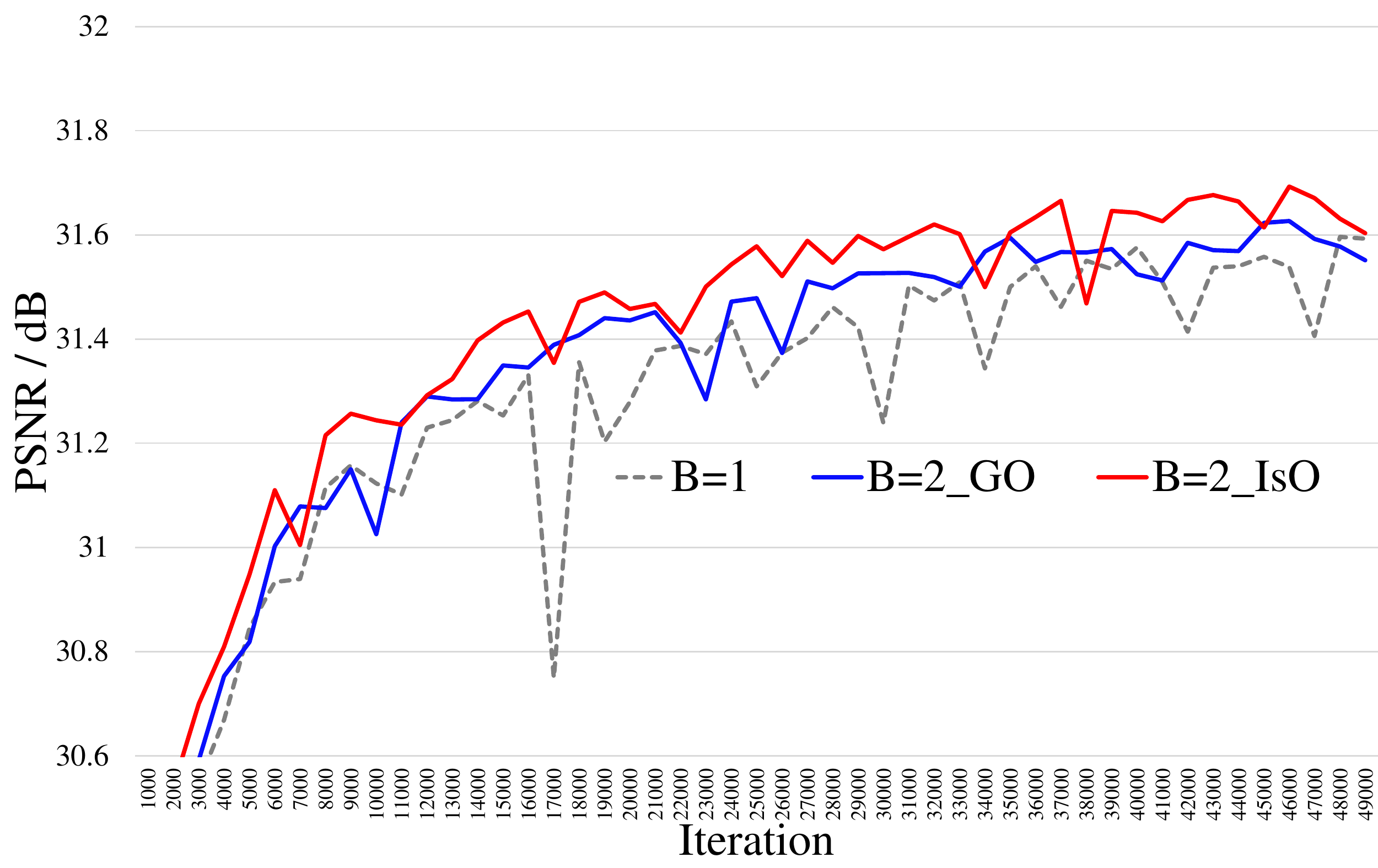}
	}\hspace{+0.4cm}
	\subfloat[B=16]{
		\footnotesize
		\includegraphics[width=0.45\linewidth]{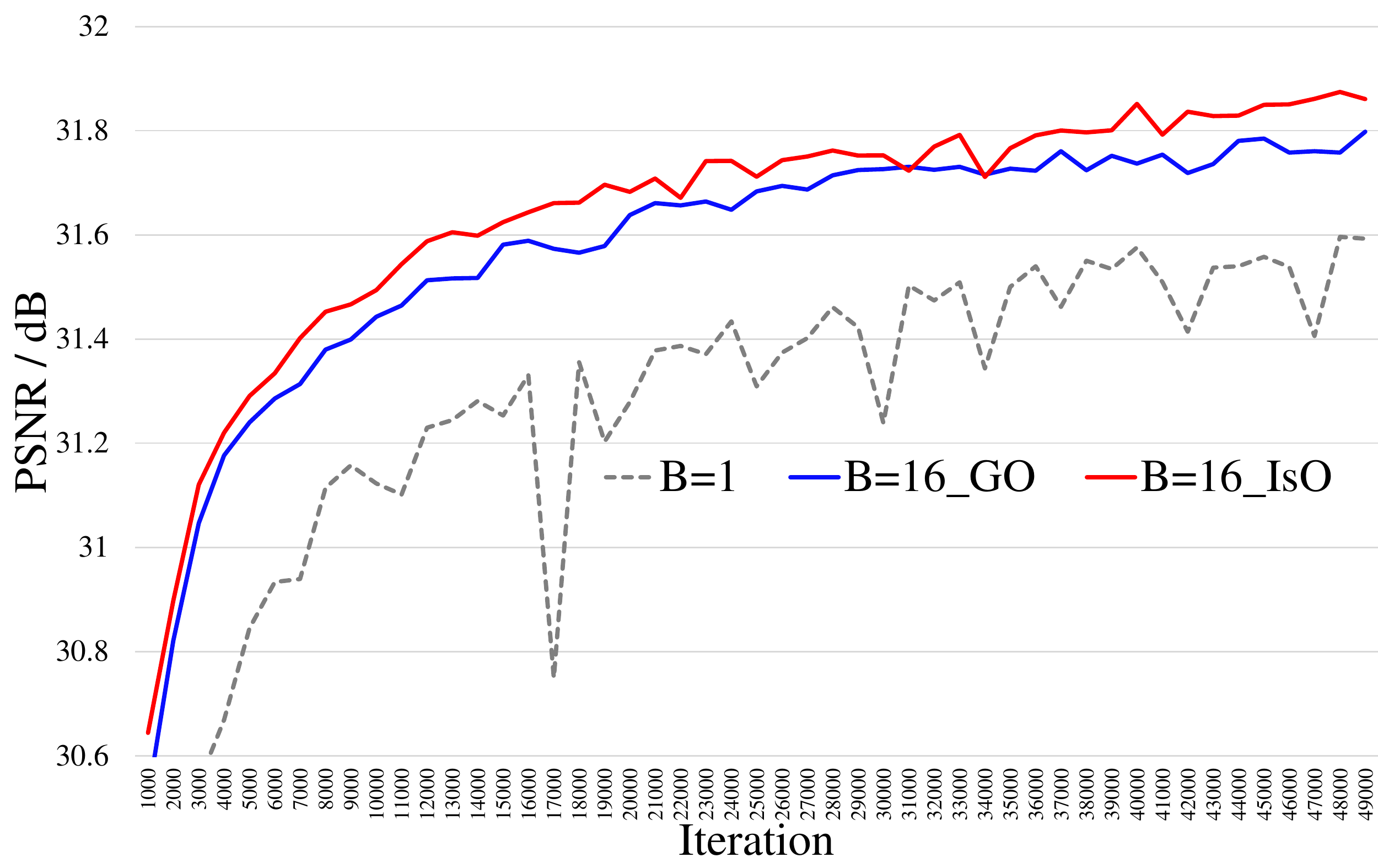}
	}
	\caption{Effect of IsO algorithm in mini-batch training: average PSNR on BSD100 dataset for $\times2$ upscaling. As batch size increases, IsO shows higher superiority to general optimization (GO).}
	\label{fig:effect_IsO}
\end{figure*}

\begin{table}
	\renewcommand\arraystretch{1.1}
	\centering
	\captionsetup{justification=centering}
	\caption{\textsc{\\Ablation study on U-HDN for $\times2$ upscaling.}}
	
	\resizebox{1\linewidth}{!}{
		\begin{tabular}{{c c|c c c c}}
			\hline
			\hline
			\multicolumn{2}{c|}{U-style residual learning}&$\times$&\checkmark&$\times$&\checkmark\\
			\multicolumn{2}{c|}{Hourglass dense block}&$\times$&$\times$&\checkmark&\checkmark\\
			\hline
			\multirow{2}{*}{Complexities}&\#Params(M)&1380.3&1380.3&1390.9&1390.9\\
			&\#FLOPs(G)&60.0&60.0&60.3&60.3\\
			\hline
			\multirow{2}{*}{Set5}&PSNR&37.50&37.61&37.54&\textbf{37.65}\\
			&SSIM&0.9584&0.9588&0.9587&\textbf{0.9590}\\
			
			\multirow{2}{*}{BSD100}&PSNR&31.80&31.84&31.82&\textbf{31.87}\\
			&SSIM&0.8950&0.8956&0.8955&\textbf{0.8960}\\
			\hline
			\hline
	\end{tabular}}
	\label{tab:ablation_study_UHDB}
\end{table}

\subsubsection{Effect of IsO algorithm}
As illustrated in Section~\ref{sec:IKM}, each image should generate its specific contextual attention to adaptively recalibrate the kernel weights. We then introduce the image-specific optimization (IsO) in mini-batch training phase as described in Algorithm 1, which is a feasible algorithm to optimize kernel weights and effectively backward the image-specific layout attentions, both of which positively guide the optimization of model. To demonstrate the effectiveness of our IsO algorithm, we conduct several experiments on optimization with different mini-batches using IsO algorithm or general optimization (GO) algorithm\footnote[1]{To keep the attentions $\boldsymbol{a}$ with same size of kernel weights $\boldsymbol{w}$, a common way is to average and repeat $\boldsymbol{a}\in\mathbb{R}^{B\times c_{in}\times \mathcal{R}}$ into $\boldsymbol{a}\in\mathbb{R}^{c_{out}\times c_{in}\times \mathcal{R}}$, then use general convolution to infer the output features as $\boldsymbol{y} = Conv(\boldsymbol{x};\boldsymbol{\hat w})$, where $\boldsymbol{\hat w}$ is the modulated kernel as in Eq.~(\ref{eq:recalibrate_kernel}).}. 

As shown in Fig.~\ref{fig:effect_IsO}, as the batch size increases, our IsO algorithm shows higher superiorities to general optimization. The main reason is that, in general optimization, when the batch size increases and the generated attentions are averaged into $\boldsymbol{a}\in\mathbb{R}^{c_{out}\times c_{in}\times \mathcal{R}}$, then the averaged attentions have lost the ability to represent the image-specific contextual information and only play a part on channel-wise recalibration as channel attention mechanism.

\subsubsection{Investigation on U-HDN}
As aforementioned in Section~\ref{sec:UHDN}, we design a new backbone U-HDN by stacking $N$=4 U-HDBs, which is a variant of dense block~\cite{HuangG2017CVPR,TongT2017ICCV} with two major amelioration: U-style residual learning and hourglass dense block.
To investigate their effect, we conduct the ablation study on the U-HDB structure and report the comparative results in TABLE~\ref{tab:ablation_study_UHDB} and find that, both of the U-style residual learning and hourglass dense block play significant roles in improving the performances.

Furthermore, from Eq.~(\ref{eq:U residual learning}), the output information flows (features) of each block have higher relevance to the head and tail dense units in the U-style residual learning, then it is better to rich these units with deeper structure, \eg, the depths in each block are set symmetrically as $[6,5,4,4,5,6]$. To demonstrate this assumption, we keep the U-style residual learning and design some variants of hourglass dense block with comparable parameters, including:
\begin{itemize}
\item ``U-HDN'' where the depthes in each block are symmetric as $[6,5,4,4,5,6]$ and the computational complexities are \#Params=1390.9K and \#FLOPs=60.3G;
\item ``U-HDN\_V1'' where the depthes in each block are symmetric as $[4,5,6,6,5,4]$ and the computational complexities are \#Params=1390.9K and \#FLOPs=60.3G;
\item ``U-HDN\_V2'' where the depthes in each block are ascending as $[2,3,4,5,6,7]$ and the computational complexities are \#Params=1401.3K and \#FLOPs=60.8G;
\item ``U-HDN\_V3'' where the depthes in each block are descending as $[7,6,5,4,3,2]$ and the computational complexities are \#Params=1401.3K and \#FLOPs=60.8G.
\end{itemize}

\begin{figure}
	\centering
	\includegraphics[width=1\linewidth]{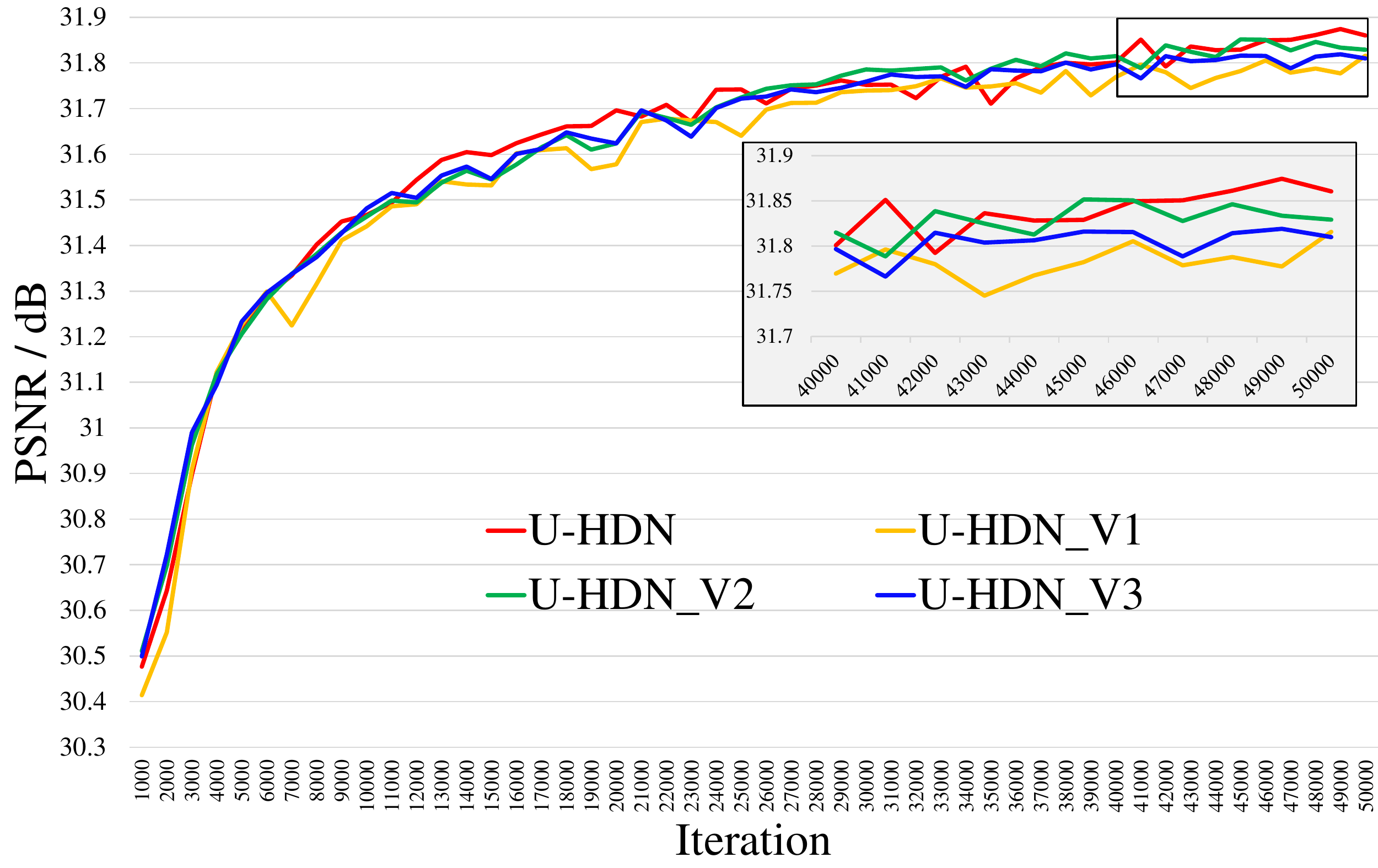}
	\caption{Effect of hourglass block learning in residual networks: average PSNR on BSD100 dataset for $\times2$ upscaling. By increasing the capacity of head or/and tail units, model achieves better performance, \ie, ``U-HDN''$>$``U-HDN\_V2''$\approx$``U-HDN\_V3''$>$``U-HDN\_V1''.}
	\label{fig:U-HDN_V}
\end{figure}
As shown in Fig.~\ref{fig:U-HDN_V}, with comparable computational complexities, U-HDN shows superiority to other variants. Hence the assumption of hourglass dense block learning that ``the desired residual features largely depend on the head and tail units'' is experimentally demonstrated, and is promising to be exploited into other applications where use dense connection. For example, by increasing the capacity of the tail unit, ``U-HDN\_V2'' and ``U-HDN\_V3'' perform better than ``U-HDN\_V1'' but worse than ``U-HDN''.

\subsection{Discussions}
As mentioned above, since the IKM is an image-specific method to adaptively modulate the convolutional kernels in deep CNNs, we should use the IsO algorithm to optimize the model in mini-batch training, and utilize the group convolutions to infer the forward propagation as illustrated in Algorithm 1. However, the existing implementation of group convolutions is less efficient than general convolution. Therefore, in spite of no additional parameters, a model with IKM runs slower than the one without it. So it is expected to exploit an efficient variant of group convolutions to accelerate the forward propagation of IKM.

Additionally, as illustrated in Eq.~(\ref{eq:reflective convolution}), the image-specific attention weights $\boldsymbol{a}$ needs a 2D matrix at same size as kernel $\boldsymbol{w}$. There also exists a large range of choosing an appropriate strategy to generate such attention weight, so it is expected to exploit more types of image-specific attention weight.

\section{Conclusions}
In this paper, to bridge the gap between the data-driven and image-specific SR methods, an image-specific convolutional kernel modulation (IKM) is proposed to adaptively modulate the convolutional kernels without any additional parameters, which achieves excellent performances against the vanilla convolution and several attention mechanisms, \eg, channel attention. Furthermore, on the optimization of IKM in mini-batch training, an image-specific optimization (IsO) algorithm is proposed and shows superiority to the general SGD optimization experimentally. Meanwhile, a U-hourglass dense network (U-HDN) is designed by utilizing U-style residual leaning and hourglass dense block learning, which are demonstrated to be appropriate architectures for utmost improving the effectiveness of IKM theoretically. Theoretical analysis and extensive experiments demonstrate the effectiveness and superiorities of the proposed methods.

\ifCLASSOPTIONcaptionsoff
  \newpage
\fi

{
	\bibliographystyle{IEEEtran}
	\bibliography{IKM}

\begin{thebibliography}{10}
\providecommand{\url}[1]{#1}
\csname url@samestyle\endcsname
\providecommand{\newblock}{\relax}
\providecommand{\bibinfo}[2]{#2}
\providecommand{\BIBentrySTDinterwordspacing}{\spaceskip=0pt\relax}
\providecommand{\BIBentryALTinterwordstretchfactor}{4}
\providecommand{\BIBentryALTinterwordspacing}{\spaceskip=\fontdimen2\font plus
\BIBentryALTinterwordstretchfactor\fontdimen3\font minus
  \fontdimen4\font\relax}
\providecommand{\BIBforeignlanguage}[2]{{%
\expandafter\ifx\csname l@#1\endcsname\relax
\typeout{** WARNING: IEEEtran.bst: No hyphenation pattern has been}%
\typeout{** loaded for the language `#1'. Using the pattern for}%
\typeout{** the default language instead.}%
\else
\language=\csname l@#1\endcsname
\fi
#2}}
\providecommand{\BIBdecl}{\relax}
\BIBdecl

\bibitem{Bicubic1981TASSP}
R.~Keys, ``Cubic convolution interpolation for digital image processing,''
  \emph{IEEE Trans. Acoustics, Speech, and Signal Process.}, vol.~29, no.~6,
  pp. 1153--1160, 1981.

\bibitem{ZhangL2006TIP}
L.~Zhang and X.~Wu, ``An edge-guided image interpolation algorithm via
  directional filtering and data fusion,'' \emph{IEEE Trans. Image Process.},
  vol.~15, no.~8, pp. 2226--2238, 2006.

\bibitem{MarquinaA2008JSC}
A.~Marquina and S.~Osher, ``Image super-resolution by tv-regularization and
  bregman iteration,'' \emph{J. Sci. Comput.}, vol.~37, no.~3, pp. 367--382,
  Dec. 2008.

\bibitem{DongW2011TIP}
W.~Dong, L.~Zhang, G.~Shi, and X.~Wu, ``Image deblurring and super-resolution
  by adaptive sparse domain selection and adaptive regularization,'' \emph{IEEE
  Trans. Image Process.}, vol.~20, no.~7, pp. 1838--1857, Jul. 2011.

\bibitem{YangJ2010TIP}
J.~Yang, J.~Wright, T.~S. Huang, and Y.~Ma, ``Image super-resolution via sparse
  representation,'' \emph{IEEE Trans. Image Process.}, vol.~19, no.~11, pp.
  2861--2873, 2010.

\bibitem{ZeydeR2010}
R.~Zeyde, M.~Elad, and M.~Protter, ``On single image scale-up using
  sparse-representations,'' in \emph{Curves and Surfaces}, 2010.

\bibitem{TimofteR2014ACCV}
R.~Timofte, V.~D. Smet, and L.~V. Gool, ``A+: Adjusted anchored neighborhood
  regression for fast super-resolution,'' in \emph{Asian Conf. Comput. Vis.},
  2014, pp. 111--126.

\bibitem{ZhuZ2014TMM}
Z.~Zhu, F.~Guo, H.~Yu, and C.~Chen, ``Fast single image super-resolution via
  self-example learning and sparse representation,'' \emph{IEEE Trans.
  Multimedia}, vol.~16, no.~8, pp. 2178--2190, 2014.

\bibitem{HuY2016TIP}
Y.~Hu, N.~Wang, D.~Tao, X.~Gao, and X.~Li, ``Serf: A simple, effective, robust,
  and fast image super-resolver from cascaded linear regression,'' \emph{IEEE
  Trans. Image Process.}, vol.~25, no.~9, pp. 4091--4102, 2016.

\bibitem{HuangY2018TIP}
Y.~Huang, J.~Li, X.~Gao, L.~He, and W.~Lu, ``Single image super-resolution via
  multiple mixture prior models,'' \emph{IEEE Trans. Image Process.}, vol.~27,
  no.~12, pp. 5904--5917, 2018.

\bibitem{Lecun1998IEEE}
Y.~Lecun, L.~Bottou, Y.~Bengio, and P.~Haffner, ``Gradient-based learning
  applied to document recognition,'' \emph{Proceedings of the IEEE}, vol.~86,
  no.~11, pp. 2278--2324, Nov 1998.

\bibitem{Krizhevsky2012NeurIPS}
A.~Krizhevsky, I.~Sutskever, and G.~E. Hinton, ``Imagenet classification with
  deep convolutional neural networks,'' in \emph{Adv. Neural Inform. Process.
  Syst.}, 2012, pp. 1097--1105.

\bibitem{Simonyan2015ICLR}
K.~Simonyan and A.~Zisserman, ``Very deep convolutional networks for
  large-scale image recognition,'' in \emph{Int. Conf. Learn. Represent.},
  2015.

\bibitem{Szegedy2015CVPR}
C.~Szegedy, W.~Liu, Y.~Jia, P.~Sermanet, S.~Reed, D.~Anguelov, D.~Erhan,
  V.~Vanhoucke, and A.~Rabinovich, ``Going deeper with convolutions,'' in
  \emph{IEEE Conf. Comput. Vis. Pattern Recog.}, 2015, pp. 1--9.

\bibitem{HeK2016CVPR}
K.~He, X.~Zhang, S.~Ren, and J.~Sun, ``Deep residual learning for image
  recognition,'' in \emph{IEEE Conf. Comput. Vis. Pattern Recog.}, 2016, pp.
  1063--6919.

\bibitem{HuangG2017CVPR}
G.~Huang, Z.~Liu, L.~van~der Maaten, and K.~Q. Weinberger, ``Densely connected
  convolutional networks,'' in \emph{IEEE Conf. Comput. Vis. Pattern Recog.},
  2017, pp. 2261--2269.

\bibitem{HuJ2018CVPR}
J.~Hu, L.~Shen, and G.~Sun, ``Squeeze-and-excitation networks,'' in \emph{IEEE
  Conf. Comput. Vis. Pattern Recog.}, 2018, pp. 7132--7141.

\bibitem{DongC2014ECCV}
C.~Dong, C.~C. Loy, K.~He, and X.~Tang, ``Learning a deep convolutional network
  for image super-resolution,'' in \emph{Eur. Conf. Comput. Vis.}, 2014, pp.
  184--199.

\bibitem{DongC2016TPAMI}
{C. Dong, C. C. Loy, K. He and X. Tang}, ``Image super-resolution using deep
  convolutional networks,'' \emph{IEEE Trans. Pattern Anal. Mach. Intell.},
  vol.~38, no.~2, pp. 295--307, 2016.

\bibitem{KimJ2016CVPR_VDSR}
J.~Kim, J.~K. Lee, and K.~M. Lee, ``Accurate image super-resolution using very
  deep convolutional networks,'' in \emph{IEEE Conf. Comput. Vis. Pattern
  Recog.}, 2016, pp. 1646--1654.

\bibitem{LedigC2017CVPR}
C.~Ledig, L.~Theis, F.~Husz{\'{a}}r, J.~Caballero, A.~P. Aitken, A.~Tejani,
  J.~Totz, Z.~Wang, and W.~Shi, ``Photo-realistic single image super-resolution
  using a generative adversarial network,'' in \emph{IEEE Conf. Comput. Vis.
  Pattern Recog.}, Jul. 2017, pp. 105--114.

\bibitem{TaiY2017CVPR}
Y.~Tai, J.~Yang, and X.~Liu, ``Image super-resolution via deep recursive
  residual network,'' in \emph{IEEE Conf. Comput. Vis. Pattern Recog.}, 2017,
  pp. 3147--3155.

\bibitem{LimB2017CVPRW}
B.~Lim, S.~Son, H.~Kim, S.~Nah, and K.~M. Lee, ``Enhanced deep residual
  networks for single image super-resolution,'' in \emph{IEEE Conf. Comput.
  Vis. Pattern Recog. Worksh.}, 2017, pp. 136--144.

\bibitem{TongT2017ICCV}
T.~Tong, G.~Li, X.~Liu, and Q.~Gao, ``Image super-resolution using dense skip
  connections,'' in \emph{Int. Conf. Comput. Vis.}, 2017, pp. 4799--4807.

\bibitem{TaiY2017ICCV}
Y.~Tai, J.~Yang, X.~Liu, and C.~Xu, ``Memnet: {A} persistent memory network for
  image restoration,'' in \emph{Int. Conf. Comput. Vis.}, 2017, pp. 4549--4557.

\bibitem{ZhangY2018CVPR}
Y.~Zhang, Y.~Tian, Y.~Kong, B.~Zhong, and Y.~Fu, ``Residual dense network for
  image super-resolution,'' in \emph{IEEE Conf. Comput. Vis. Pattern Recog.},
  2018, pp. 2472--2481.

\bibitem{ZhangY2021TPAMI}
\protect{Y. Zhang}, Y.~Tian, Y.~Kong, B.~Zhong, and Y.~Fu, ``Residual dense
  network for image restoration,'' \emph{IEEE Trans. Pattern Anal. Mach.
  Intell.}, vol.~43, no.~7, pp. 2480--2495, 2021.

\bibitem{ZhangX2021TMM}
X.~Zhang, P.~Gao, S.~Liu, K.~Zhao, G.~Li, L.~Yin, and C.~W. Chen, ``Accurate
  and efficient image super-resolution via global-local adjusting dense
  network,'' \emph{IEEE Trans. Multimedia}, vol.~23, pp. 1924--1937, 2021.

\bibitem{DongC2016ECCV}
C.~Dong, C.~C. Loy, and X.~Tang, ``Accelerating the super-resolution
  convolutional neural network,'' in \emph{Eur. Conf. Comput. Vis.}, 2016, pp.
  391--407.

\bibitem{ShiW2016CVPR}
W.~Shi, J.~Caballero, F.~Husz{\'{a}}r, J.~Totz, A.~P. Aitken, R.~Bishop,
  D.~Rueckert, and Z.~Wang, ``Real-time single image and video super-resolution
  using an efficient sub-pixel convolutional neural network,'' in \emph{IEEE
  Conf. Comput. Vis. Pattern Recog.}, 2016, pp. 1874--1883.

\bibitem{ZhangY2018ECCV}
Y.~Zhang, K.~Li, K.~Li, L.~Wang, B.~Zhong, and Y.~Fu, ``Image super-resolution
  using very deep residual channel attention networks,'' in \emph{Eur. Conf.
  Comput. Vis.}, 2018, pp. 294--310.

\bibitem{WooS2018ECCV}
S.~Woo, J.~Park, J.-Y. Lee, and I.~S. Kweon, ``Cbam: Convolutional block
  attention module,'' in \emph{Eur. Conf. Comput. Vis.}, 2018, pp. 3--19.

\bibitem{HuY2020TCSVT}
Y.~Hu, J.~Li, Y.~Huang, and X.~Gao, ``Channel-wise and spatial feature
  modulation network for single image super-resolution,'' \emph{IEEE Trans.
  Circuit Syst. Video Technol.}, vol.~30, no.~11, pp. 3911--3927, 2020.

\bibitem{WangXL2018CVPR}
X.~Wang, K.~Yu, C.~Dong, and C.~C. Loy, ``Recovering realistic texture in image
  super-resolution by deep spatial feature transform,'' in \emph{IEEE Conf.
  Comput. Vis. Pattern Recog.}, 2018, pp. 7794--7803.

\bibitem{ZhangY2019ICLR}
\protect{Y. Zhang}, K.~Li, K.~Li, L.~Wang, B.~Zhong, and Y.~Fu, ``Residual
  non-local attention networks for image restoration,'' in \emph{Int. Conf.
  Learn. Represent.}, 2019.

\bibitem{HuangY2021TIP}
Y.~Huang, J.~Li, X.~Gao, Y.~Hu, and W.~Lu, ``Interpretable detail-fidelity
  attention network for single image super-resolution,'' \emph{IEEE Trans.
  Image Process.}, vol.~30, pp. 2325--2339, 2021.

\bibitem{ShocherA2018CVPR}
A.~Shocher, N.~Cohen, and M.~Irani, ``Zero-shot super-resolution using deep
  internal learning,'' in \emph{IEEE Conf. Comput. Vis. Pattern Recog.}, 2018,
  pp. 3118--3126.

\bibitem{UlyanovD2018CVPR}
D.~Ulyanov, A.~Vedaldi, and V.~Lempitsky, ``Deep image prior,'' in \emph{IEEE
  Conf. Comput. Vis. Pattern Recog.}, 2017, pp. 9446--9454.

\bibitem{ChenL2017CVPR}
L.~Chen, H.~Zhang, J.~Xiao, L.~Nie, J.~Shao, and T.-S. Chua, ``Sca-cnn: Spatial
  and channel-wise attention in convolutional networks for image captioning,''
  in \emph{IEEE Conf. Comput. Vis. Pattern Recog.}, 2017, pp. 6298--6306.

\bibitem{XuK2015ICML}
K.~Xu, J.~Ba, R.~Kiros, A.~Courville, R.~Salakhutdinov, R.~Zemel, and
  Y.~Bengio, ``Show, attend and tell: Neural image caption generation with
  visual attention,'' in \emph{Int. Conf. Mach. Learn.}, 2015, pp. 2048--2057.

\bibitem{WangXT2018CVPR}
X.~Wang, K.~Yu, C.~Dong, and C.~C. Loy, ``Recovering realistic texture in image
  super-resolution by deep spatial feature transform,'' in \emph{IEEE Conf.
  Comput. Vis. Pattern Recog.}, 2018, pp. 606--615.

\bibitem{JetleyS2018ICLR}
S.~Jetley, N.~A. Lord, N.~Lee, and P.~H.~S. Torr, ``Learn to pay attention,''
  in \emph{Int. Conf. Learn. Represent.}, 2018.

\bibitem{ZhangJ2018IJCV}
J.~Zhang, S.~A. Bargal, Z.~Lin, J.~Brandt, X.~Shen, and S.~Sclaroff, ``Top-down
  neural attention by excitation backprop,'' \emph{Int. J. Comput. Vis.}, vol.
  2018, no. 126, pp. 1084--1102, Dec 2018.

\bibitem{KimJ2016CVPR_DRCN}
J.~Kim, J.~K. Lee, and K.~M. Lee, ``Deeply-recursive convolutional network for
  image super-resolution,'' in \emph{IEEE Conf. Comput. Vis. Pattern Recog.},
  2016, pp. 1637--1645.

\bibitem{HuiZ2018CVPR}
Z.~Hui, X.~Wang, and X.~Gao, ``Fast and accurate single image super-resolution
  via information distillation network,'' in \emph{IEEE Conf. Comput. Vis.
  Pattern Recog.}, 2018, pp. 723--731.

\bibitem{LiJ2018ECCV}
J.~Li, F.~Fang, K.~Mei, and G.~Zhang, ``Multi-scale residual network for image
  super-resolution,'' in \emph{Eur. Conf. Comput. Vis.}, 2018, pp. 527--542.

\bibitem{AhnN2018ECCV}
N.~Ahn, B.~Kang, and K.-A. Sohn, ``Fast, accurate, and lightweight
  super-resolution with cascading residual network,'' in \emph{Eur. Conf.
  Comput. Vis.}, 2018, pp. 256--272.

\bibitem{FengR2020ECCV}
R.~Feng, W.~Guan, Y.~Qiao, and C.~Dong, ``Exploring multi-scale feature
  propagation and communication for image super resolution,'' \emph{Eur. Conf.
  Comput. Vis.}, 2020.

\bibitem{LaiW2019TPAMI}
W.-S. Lai, J.-B. Huang, N.~Ahuja, and M.-H. Yang, ``Fast and accurate image
  super-resolution with deep laplacian pyramid networks,'' \emph{IEEE Trans.
  Pattern Anal. Mach. Intell.}, vol.~41, no.~11, pp. 2599--2613, 2019.

\bibitem{ZhangD2021TMM}
D.~Zhang, J.~Shao, Z.~Liang, L.~Gao, and H.~T. Shen, ``Large factor image
  super-resolution with cascaded convolutional neural networks,'' \emph{IEEE
  Trans. Multimedia}, vol.~23, pp. 2172--2184, 2021.

\bibitem{HuiZ2019ACMMM}
Z.~Hui, X.~Gao, Y.~Yang, and X.~Wang, ``Lightweight image super-resolution with
  information multi-distillation network,'' in \emph{ACM Int. Conf.
  Multimedia}, 2019, pp. 2024--2032.

\bibitem{TianC2021TMM}
C.~Tian, Y.~Xu, W.~Zuo, B.~Zhang, L.~Fei, and C.-W. Lin, ``Coarse-to-fine cnn
  for image super-resolution,'' \emph{IEEE Trans. Multimedia}, vol.~23, pp.
  1489--1502, 2021.

\bibitem{BevilacC2012BMVC}
M.~Bevilacqua, A.~Roumy, C.~Guillemot, and M.-L.~A. Morel, ``Low-complexity
  single-image super-resolution based on nonnegative neighbor embedding,'' in
  \emph{Brit. Mach. Vis. Conf.}, 2012.

\bibitem{ArbelaezP2011TPAMI}
P.~Arbel{\'{a}}ez, M.~Maire, C.~C. Fowlkes, and J.~Malik, ``Contour detection
  and hierarchical image segmentation,'' \emph{IEEE Trans. Pattern Anal. Mach.
  Intell.}, vol.~33, no.~5, pp. 898--916, 2011.

\bibitem{TimofteR2017CVPRW}
R.~Timofte, E.~Agustsson, L.~V. Gool, M.-H. Yang, L.~Zhang, and et~al, ``Ntire
  2017 challenge on single image super-resolution: Methods and results,'' in
  \emph{IEEE Conf. Comput. Vis. Pattern Recog. Worksh.}, 2017, pp. 1110--1121.

\bibitem{KingmaD2014ICLR}
D.~P. Kingma and J.~Ba, ``Adam: {A} method for stochastic optimization,'' in
  \emph{Int. Conf. Learn. Represent.}, 2014.

\bibitem{HeX2019CVPR}
X.~He, Z.~Mo, P.~Wang, Y.~Liu, M.~Yang, and J.~Cheng, ``Ode-inspired network
  design for single image super-resolution,'' in \emph{IEEE Conf. Comput. Vis.
  Pattern Recog.}, 2019, pp. 1732--1741.

\bibitem{YuF2016ICLR}
F.~Yu and V.~Koltun, ``Multi-scale context aggregation by dilated
  convolutions,'' in \emph{Int. Conf. Learn. Represent.}, 2016.

\end{thebibliography}
}

\end{document}